\documentclass[prd,twocolumn,floats,superscriptaddress,eqsecnum,floatfix,
showpacs]{revtex4-1}

\usepackage{amssymb,amsmath}
\usepackage{epsfig}

\pdfoutput=1
\usepackage{amssymb,amsmath}
\usepackage{epsfig}
\usepackage{color}
\usepackage{datetime}
\usepackage[utf8]{inputenc}

\newcommand{\ve}{\varepsilon}

\newcommand{\pa}{\partial}

\newcommand{\la}{\langle}
\newcommand{\ra}{\rangle}

\newcommand{\be}{\begin{equation}}
\newcommand{\ee}{\end{equation}}
\newcommand{\ba}{\begin{eqnarray}}
\newcommand{\ea}{\end{eqnarray}}
\newcommand{\beg}{\begin{gather*}}
\newcommand{\eng}{\end{gather*}}

\newcommand{\hh}{,\hspace{0.5cm}}
\newcommand{\hhh}{,\hspace{0.2cm}}

\newcommand{\eq}[1]{(\ref{#1})}

\newcommand{\n}[1]{\label{#1}}

\newcommand{\ins}[1]{{\mbox{\tiny #1}}}

\newcommand{\const}{\mbox{const}}

\newcommand{\ts}[1]{{\boldsymbol{#1}}}

\def\XXint#1#2#3{{\setbox0=\hbox{$#1{#2#3}{\int}$ }
\vcenter{\hbox{$#2#3$ }}\kern-.6\wd0}}

\begin{document}

\title{Quantum radiation from an evaporating non-singular  black hole}

\author{Valeri P. Frolov}
\email{vfrolov@ualberta.ca}
\affiliation{Theoretical Physics Institute, Department of Physics\\
University of Alberta, Edmonton, Alberta, Canada T6G 2E1}

\author{Andrei Zelnikov}
\email{zelnikov@ualberta.ca}
\affiliation{Theoretical Physics Institute, Department of Physics\\
University of Alberta, Edmonton, Alberta, Canada T6G 2E1}

\begin{abstract}
In this paper we study quantum radiation from an evaporating spherically symmetric non-singular black hole. We used a modified Hayward metric for a description of a non-singular black hole interior. We assume that the mass parameter of this metric depends on the advanced time, and choose this dependence so that it properly reproduces both black hole formation and its subsequent evaporation. We consider a quantum massless scalar field propagating in this geometry and use 2D approximation for the calculation of the quantum average of the stress-energy tensor in the initial vacuum state. For the calculation of this quantity it is sufficient to find a map between the Killing times $u_+$ and $u_-$ at the future and past null infinities, established by the propagation of the radial null rays.
In this formalism the quantum energy flux at the future null infinity can be expressed in terms of the function $u_+(u_-)$ and its derivatives up to the third order. We developed a special formalism, which allows one to reduce the problem of the calculation of the quantum energy flux and other observables to a solution of a simple set of ordinary differential equations. We used this approach to study quantum effects in two cases: i) with the trivial, $\alpha=1$ and ii) the non-trivial, $\alpha\ne1$, redshift function. We demonstrated, that in both cases there exists an outburst of the quantum energy radiation from the inner domain of the black hole, close to the inner part of its apparent horizon. For $\alpha=1$ this outburst is exponentially large. Its appearance is a direct consequence of the so-called mass inflation effect. We also demonstrated, that this severe problem can be solved by a proper choice of the redshift function. However, even in this case the emitted energy can be much larger than the initial mass of the evaporating black hole. This means that for a construction of a self-consistent model of a non-singular evaporating black hole the back-reaction effects are highly important.


\end{abstract}


\maketitle


\section{Introduction}

A long-standing problem of the general relativity is the inevitable existence of singularities both in cosmology and in black holes. It is generally believed that a proper theory of quantum gravity would solve this problem. There exist many publications which support this point of view.  One of the options is to modify the Einstein equations in the spacetime domains where the curvature becomes high. Such a modification may have different origins. It may be connected with the necessity to include the back-reaction of the vacuum polarization and particle creation in the presence of quantum fields, or/and as a result of the quantization of the gravitational field itself. More deep reason might be that the gravity is an emergent phenomenon, so that the effective gravity equations arise as the result of averaging over the degrees of freedom of the constituents of the background fundamental theory, such as the string theory or the loop gravity. Another option is to adopt a special form of the modified gravity equations and to consider them as a new fundamental law of gravity. Such an approach is widely used in the cosmology, both to describe the evolution of the early universe as well as its present-time acceleration. One of the interesting  recent example of the modified gravity is a so called ghostfree gravity \cite{Tomboulis:1997gg, Biswas:2011ar, Modesto:2011kw, Modesto:2012ys, Biswas:2013cha, Biswas:2013kla,
Modesto:2014lga, Talaganis:2014ida,Tomboulis:2015gfa,Tomboulis:2015esa}.

In the absence of the reliable fundamental theory, describing physics in the high-curvature regime, one can also use a more naive, phenomenological approach. In such an approach, one usually assumes that the gravity in the high curvature regime is still described by a classical metric. There exist a fundamental energy scale parameter $\mu$, and related to it the fundamental length scale $\ell=\mu^{-1}$. When the curvature becomes comparable with $\ell^{-2}$ a solution of the classical Einstein equations should be modified. One does not fix the gravity equations, but instead of this one imposes a number of ``natural'' restrictions on  the form of the metric. To be more concrete we assume that: (i) The corrections to the solution of the Einstein equations are small in the domain of small curvature, ${\cal R}\ll \ell^{-2}$; (ii) The metric is regular; (iii) The curvature obeys the limiting curvature condition, that is its value is uniformly restricted by some fundamental value, $|{\cal R}|\le C \ell^{-2}$ \cite{Markov:1982,Markov:1984ii} (see also \cite{Polchinski:1989ae}). These assumptions require explanations. We did not specify the meaning of ${\cal R}$. In fact one might think that
${\cal R}$ is any of scalar invariants constructed from the Riemann curvature and its covariant derivatives, which has the proper dimensionality of $\ell^{-2}$.  In practice, one usually considers the invariants $|R|$, $\sqrt{ | R_{\mu\nu} R^{\mu\nu} | }$ and
$\sqrt{ | R_{\mu\nu\alpha\beta} R^{\mu\nu\alpha\beta} | }$. It should be emphasized, that
the dimensionless constant $C$, which  depends on the choice of the invariant, is determined only by the  theory and it cannot depend on the particular choice of a solution and its parameters.

Black-hole metrics, obeying the above conditions, are known as non-singular ones. There are many publications discussing properties of different models of non-singular evaporating black holes (see e.g. \cite{Frolov:1981mz,Roman:1983zza,Borde:1996df,Hayward:2005gi,Hossenfelder:2009fc,Bambi:2013gva,Bambi:2013caa, Hawking:2014tga, Frolov:2014jva,Bardeen:2014uaa,Haggard:2014rza, Barrau:2015uca,Haggard:2015iya}).

In this paper we continue study quantum effects in the background on a non-singular evaporating black hole. Already in the first publication on this subject \cite{Bolashenko:1986mr} it was demonstrated that in the presence of the inner horizon, which exists in the non-singular black holes, the quantum effects results in the huge outburst of the quantum radiation from the black-hole interior. This result was confirmed later in \cite{Lorenzo,Frolov:2016gwl}.
Total energy radiated in this outburst can be estimated as $\mu \exp(q/\ell)$, where $q$ is the time of existence of the black hole. There is a direct relation of this result with a well known phenomenon of mass inflation \cite{Brown:2011tv}. It was demonstrated in the paper \cite{Frolov:2016gwl} that this is a consequence of the large value of the (negative) surface gravity on the inner horizon, which in a standard (Hayward-type \cite{Hayward:2005gi} models) is of the order of $\ell^{-1}$. The surface gravity on the inner horizon can be suppressed by including a specially chosen redshift factor in the metric (see e.g. \cite{Lorenzo,Frolov:2016pav}). In our previous paper \cite{Frolov:2016gwl}, we demonstrated that the exponentially large quantum energy release from the interior of a nonsingular black-hole is really suppressed by a proper choice of the redshift factor. However, this energy still remains large and can easily exceed the initial mass of the black hole. This means that there still exist a problem of consistency of the corresponding non-singular black-hole model \cite{Bolashenko:1986mr}.

The results presented in \cite{Frolov:2016gwl} are restricted by a special so called sandwich model. In this model  it is assumed that a black hole is created by the collapse of a spherical null shell of mass $M$, and its existence is terminated by a subsequent collapse of the other spherical null shell of mass $-M$. Such a model has only two parameters, the mass $M$ and duration of the black hole existence $q$. This makes the analysis of the results in different regimes quite easy. At the same time, certainly this model is quite artificial. The goal of the present paper is to perform calculations of the quantum energy flux from a more realistic evaporating spherically symmetric nonsingular black hole.

Radial null rays, propagating in a nonsingular black-hole geometry, establish relation between the retarded time $u_+$ of the arrival to ${\cal I}^+$ and the advanced time $u_-$ when the ray left ${\cal I}^-$. In the adopted 2D approximation the energy flux of a quantum massless scalar field can be expressed in terms of the function $u_-=u_-(u_+)$ and its derivatives up to the third order. Reuter \cite{Reuter:1988nt} demonstrated that this quantity is, in fact, proportional to the so called {\em Schwarzian derivative}. Based on this result, we show that this and other similar objects, describing observables at ${\cal I}^+$ for an evaporating black hole can be obtained by integrating a set of rather simple ordinary differential equations. To prove this result we use special properties of the congruence (beam) of outgoing null rays. The first part of the paper is devoted to derivation of these results, which we believe are new. In the second part we present the results of the numerical calculations for the characteristics of radiation from the evaporating nonsingular black hole. Summary of the obtained results and their discussion conclude the paper.

\section{Non-singular models of evaporating black holes}

\subsection{Spherically symmetric regular black holes}

The most general spherically symmetric metric in the four dimensional spacetime can be written in the form
\be\begin{split}\label{n1}
&dS^2=\sigma^2ds^2,\\
&ds^2=-\alpha^2 f dv^2 +2\alpha dv dr +r^2 d\omega^2.
\end{split}\ee
We study quantum effects on the background of regular black holes, that are solutions of some modification of the gravity theory. This gravity theory is assumed to reproduce the Einstein equations at large scales but remove singularities at small distances. We denote by  $\ell$ the critical length  at which deformation of the Einstein equations becomes large.
This critical length $\ell$ is a characteristic of the gravity theory and can be of the order of the Planck length $l_{Planck}$ or much larger. In the latter case quantum fluctuations of the metric in the domain between $\ell$  and $l_{Planck}$ are still small while the black hole solutions may be free of singularities. In the presence of critical length parameter it is natural to choose the scale factor  $\sigma$  of the dimensionality of
$[ length]$ to coincide with $\ell$. $ds^2$ is the dimensionless metric and coordinates $v$ and $r$ are also dimensionless.
In a general case, the metric coefficients $f$ and $\alpha$ are functions of both variables, $v$ and $r$. It is easy to check that
\be
f=g^{\mu\nu}r_{,\mu} r_{,\nu}\, .
\ee
Points where $f=0$ form an apparent horizon. For regular black holes the apparent horizon typically consists of an outer $r_\ins{1}(v)$ and an inner $r_\ins{2}(v)$ branches. In the static case the apparent horizons coincide with the Killing horizons and one can calculate their surface gravities
\be\n{kappa12}
\kappa_\ins{1,2}=\left({\alpha\over 2}{\partial f\over \partial r} \right)\Big|_{r=r_\ins{1,2}}.
\ee

We assume that the spacetime is asymptotically flat so that
\be
f(v,r)|_{r\to\infty}=1\, .
\ee
We call $\alpha(v,r)$ a red-shift function. Using an ambiguity in the choice of $v$, we put $\alpha (v,r)|_{r\to\infty}=1$. In a static spacetime with a Killing vector $\ts{\xi}$ one has
\be
\ts{\xi}=-\alpha^2 f\, .
\ee

In what follows we assume that the metric (\ref{n1}) is regular at the center $r=0$, that is the curvature invariants are finite there. It is easy to show that this regularity condition implies
\ba
f&=&1+{1\over 2}f_2(v) r^2+\ldots\, ,\n{ff0}\\
\alpha &=&\alpha_0(v)[1+{1\over 2}\alpha_2(v) r^2]+\ldots\,  .\n{aa0}
\ea
In a general case, when $\alpha_0(v)\ne 1$, the rate of  the proper time $\tau$ at the center differs from the rate of time $v$
\be
d\tau=\alpha_0(v) dv\, .
\ee

\subsection{Propagation of null rays}

\begin{figure}[tbp]
\centering
\includegraphics[width=6.5cm]{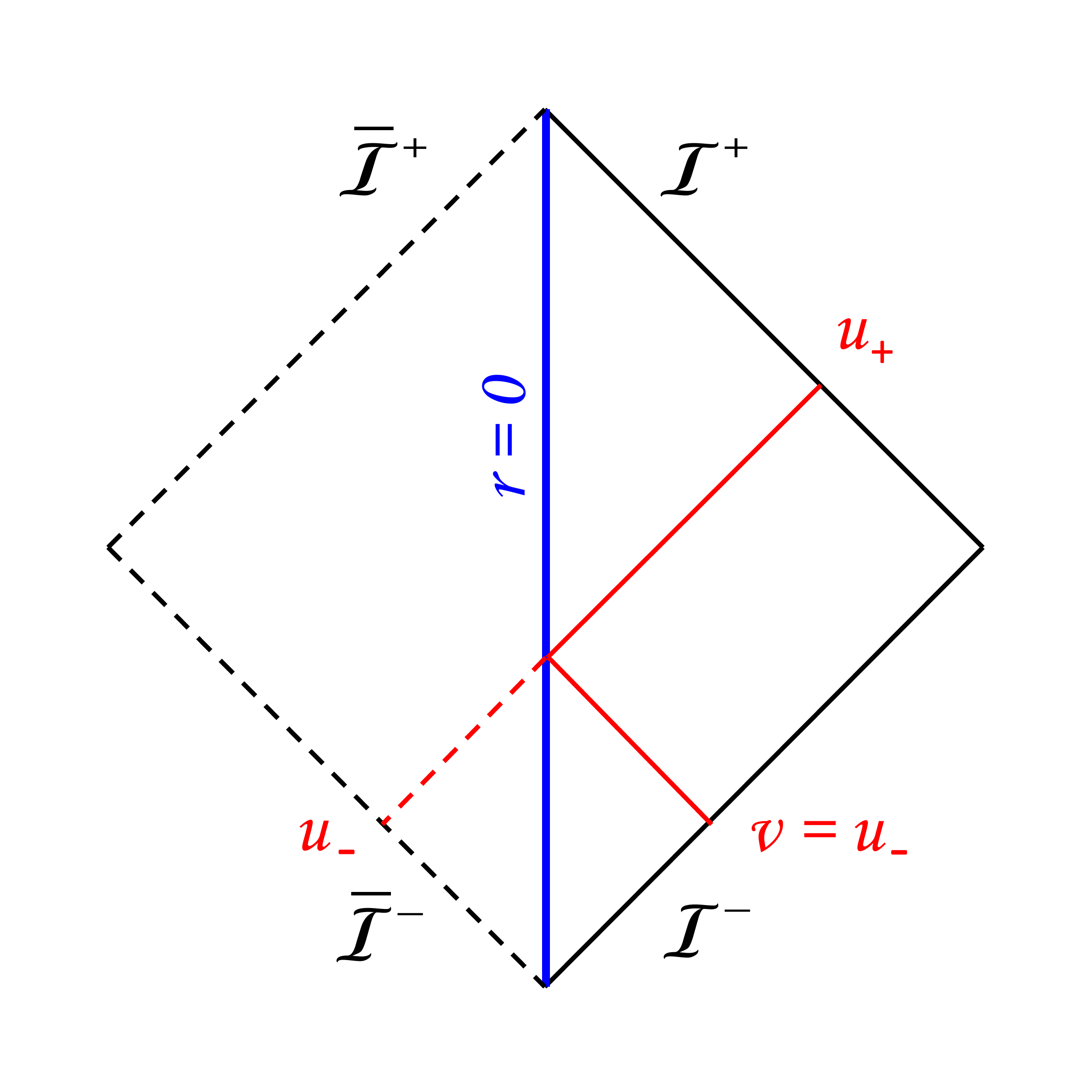}
  \caption{
  Propagation of radial null rays in an asymptotically flat spacetime.
\label{Fig1}}
\end{figure}

In the chosen coordinates, radial incoming null rays are described by the relation $v=$const , while the outgoing null rays satisfy the following equation
\be\n{outr}
{dr\over dv}={\cal Z}(v,r)\hh {\cal Z}={1\over 2} \alpha f\, .
\ee
Consider an outgoing radial null ray which reaches the future null infinity ${\cal I}^+$ at the retarded time $u=u_+$ and trace it backward in time. We denote by $u_-$ time $v$ when it crosses the center $r=0$. After this it propagates to the past null infinity ${\cal I}^-$ at the same advanced time $u_-$. The relation $u_-=u_-(u_+)$ establishes a map between ${\cal I}^+$ and ${\cal I}^-$. An adopted notation $u_-$ for the coordinate on ${\cal I}^-$ might look misleading, so that it requires some explanations. There are two equivalent ways how one can describe the propagation of the null rays in the spherical geometry (see Figure~\ref{Fig1}). One can consider either a complete two-plane representing $(t,r)$ sector of the spacetime, or only a half of it, restricted by the ``boundary'' $r=0$. In the former case, there are two copies of ${\cal I}^+$ and ${\cal I}^-$, that are sections of the complete $(t,r)$ two plane of the surfaces of the future and past infinities. A null ray, which reaches the ``right'' future null infinity at the moment of the retarded time $u_+$ is emitted from the ``left'' copy of the past null infinity at the moment of the retarded time $u_-$. If one uses only the half of the $(t,r)$ plane, the same motion is represented by a line, which is reflected at $r=0$. Since the tangent vector to the null ray trajectory is uniquely determined by equation (\ref{outr}), any two null rays cannot intersect. This means that $u_+(u_-)$ is a monotonically increasing function, so that $du_+/du_->0$.

\subsection{Quantum fluxes at ${\cal I}^+$}

In what follows, we study quantum radiation of a massless scalar fields in the geometry of evaporating regular``black holes''. Using  decomposition of the quantum modes in spherical harmonics one can show  \cite{Frolov:1998wf} that the main contribution to the Hawking radiation comes from $S$-modes. Some exact properties of quantum radiation in $S$-mode are possible to derive \cite{Mukhanov:1994ax} but its complete analysis is still quite difficult, because of coupling of  $2D$  conformal matter fields to a dilaton.
If we neglect scattering of  $S$-modes by the gravitational potential, then the theory is effectively reduced to the quantum theory of a $2D$ conformal scalar field.
One can expect that this simplified $2D$ approach is good enough to describe  qualitative features of quantum radiation.

We start with an action for a two-dimensional  conformal scalar field
\be
S=-{1\over 2}\int d^2x\sqrt{-g}\,(\nabla\hat{\varphi})^2,
\ee
where the two-dimensional metric is given by \eq{n1}.

The density of an energy flux of quantum radiation it is proportional to the Planck constant and can be written as
\be
T^{vv}=(\ell_{Pl}/\ell)^2 {\cal E}\, ,
\ee
where ${\cal E}$ is the dimensionless rate of the energy emission in the adopted $\ell$ units.
The energy flux of massless particles, created from the initial vacuum state, is given by the following expression
\be\n{EEEE}
{\cal E}={1\over 48\pi}\left[ -2{d^2 P\over du^2}+\left({d P\over du}\right)^2\right]\, ,
\ee
where
\be\label{P}
P=\ln \left| {du_-\over du_+}\right|\, .
\ee
This relation directly follows from a general result obtained by Fulling and Christensen \cite{Christensen:1977jc} for the quantum average of the stress-energy tensor of a massless scalar field in two dimensions, reconstructed from the conformal anomaly. The same expression can be also obtained by the variation of the Polyakov effective action with respect to the metric \cite{Frolov:1984,Vilkovisky:1985}.

\subsection{Schwarzian derivative and its ``relatives''}

As we already mentioned, radial null geodesics establish a map between ${\cal I}^-$ and ${\cal I}^+$. In the 2D approximation the classical and quantum characteristics of an evaporating black hole can be found if this map is known. To describe this map one can use either the function $u_+=u_+(u_-)$, or its inverse $u_-=u_-(u_+)$. The energy flux from an evaporating black hole can be naturally expressed in terms of so called {\it Schwarzian derivatives} \cite{Reuter:1988nt}. Since we shall be using this formalism in our further calculations let us  briefly describe it.

Consider a function of one variable $y=y(x)$ and let $x=x(y)$ be its inverse, that is $x(y(x))=x$. Their derivatives are connected as follows
\be
y'={1\over x'}\, .
\ee
For briefness, we denote by prime a derivative of a function of one variable with respect to its argument.
It is convenient to introduce the following notations
\ba
&&[y,x]=\ln |y'|\, ,\\
&&\la y,x\ra ={y''\over y'}\, ,\\
&& \{ y,x\}={y'''\over y'}-{3\over 2} \left({y''\over y'}\right)^2\, .
\ea
The last expression is known as the {\it Schwarzian derivative}, or simply the {\it Schwarzian}. These objects have special properties, that make them useful in the further consideration.

For the inverse functions $y(x)$ and $x(y)$ one has
\ba
&&[y,x]=-[x,y]\, ,\\
&& \la y,x\ra = -y' \la x,y\ra \, ,\n{yx}\\
&&\{y,x\}=-(y')^2 \{x,y\}\, .\n{yyxx}\\
&&
\ea
The objects $[y,x]$, $\la y,x\ra$ and $\{y,x\}$ satisfy simple chain rules, which will be used in our calculations. Namely, let $f\circ g$ is a function defined as
\be
f\circ g(z)=f(g(z))\, .
\ee
Then one has
\ba
&& [f\circ g,z] = \left. [ f, g]\right|_{g=g(z)} + [ g, z]\, ,\\
&& \la f\circ g,z\ra = \left.\la f, g\ra\right|_{g=g(z)} g'(z) + \la g, z\ra\, ,\\
&& \{ f \circ g, z\}=\left. \{ f, g \} \right|_{g=g(z)} (g'(z))^2 +\{ g, z\}\, .
\ea

\subsection{Observables}

Suppose the map function $u_-(u_+)$ is known, then one can use it to calculate special observables. For example, a gain function $\beta$, which is the ratio the energy (frequency) of the outgoing photon to its original in-falling energy (frequency), is given by the expression
\be
\beta={du_-\over du_+}\, .
\ee
The logarithm of $\beta$
\be\label{P1}
P=\ln \beta=[u_-,u_+]\, ,
\ee
is used sometime as a measure of the radiation entropy density (see \cite{Page:1993wv,Bianchi:2014bma})
\be
S(v)=-{1\over 12} P\, .
\ee
The function $W$, which characterizes the density of outgoing trajectories, is
\be\label{W1}
W=\la u_-,u_+\ra\, .
\ee
The energy flux \eq{EEEE} is given by the Schwarzian derivative \cite{Reuter:1988nt}
\be\n{EESa}
{\cal E}=-{1\over 24\pi} \{u_-,u_+\}\, .
\ee

The same observables can be easily written in terms of the inverse function $_+(u_-)$. Namely
\ba
&&P=-[u_+,u_-] ,\n{P2}\\
&&W=- e^P \la u_+,u_-\ra ,\n{W2}\\
&&{\cal E}={1\over 24\pi}e^{2P} \{u_+,u_-\}\n{E2}.
\ea

In what follows, we describe  two methods of calculations of the observables, such as the gain function and quantum energy flux at ${\cal I}^+$. These methods differs by the choice of the function, representing the map, as a starting point of the calculations. For the calculation of $u_+=u_+(u_-)$ and its required derivatives one uses the evolution of the radial null ray from the past to the future. We call this method a {\it bottom-up approach}. If one uses the function $u_-=u_-(u_+)$ and its derivatives one needs to consider the evolution of the radial null ray backward in time. We call this method a {\it top-down approach}. Certainly,  both methods of the calculations should give the same result for any observable. This property provides one with a useful test of the numerical calculations.

\section{Beams of null rays}

\subsection{Beam equations}

We now demonstrate, that the calculation of such observables as $\beta$, $P$, $W$ and ${\cal E}$ can be effectively reduced to solving a rather simple set of the ordinary differential equations.

For this purpose we consider a beam of the out-going null rays and use a parameter $z$ to ``enumerate'' them. This means that the beam is described by a function $r(v,z)$ and the parameter $z$ is constant at each ray of the beam. We choose one of the null rays as {\it fiducial} and put $z=0$ for this ray
\be
r_0(v)=r(v,z=0)\, .
\ee
The function $r(v,z)$ obeys the equation
\be
\dot{r}={\cal Z}(v,r(v,z))\, .
\ee

For the rays close to the fiducial ray one has the following decomposition
\ba
r(v,z)&=&r_0(v)+\Delta r(v,z)\, ,\n{rr}\\
\Delta r(v,z)&=&\sum_{n=1} {z^n\over n!} r_n(v) \, .\n{rrx}
\ea
In the right-hand side of the ray propagation equation (\ref{outr}) one has ${\cal Z}(v,r)$. It is convenient to write it in the form
\ba
&&Z(v,z)\equiv {\cal Z}(v,r(v,z))=Z_0(v)+\Delta Z(v,z)\, ,\n{ZZ}\\
&&\Delta Z(v,z)=\sum_{n=1} {z^n\over n!} Z_n(v) \, .\n{ZZx}
\ea
If the  function $Z(v,z)$ is known, the equation (\ref{outr}) reduces to the following simple set of the equations
\be\n{rrr}
\dot{r}_0=Z_0(v)\hh \dot{r}_n=Z_n(v)\, .
\ee
Here and later a dot denotes a derivative along the fiducial null ray, that is
\be
\dot{A}=\left. \pa_v A(v,z)\right|_{z=0}\, .
\ee
In practice we need to know equations from the system (\ref{rrr}) only up to some order $n$. We call this set of equation a {\em truncated system of the order $n$}.

In order to specify a solution, one needs to impose an initial data. For this purpose we chose a null surface $v=v_0$ and put $r(v_0,z)=r_0+z$, where $r_0$ is the radius of the point where the fiducial ray crosses the chosen surface. The functions $r_n(v)$ obey the equations (\ref{rrr}) with the initial conditions $r_0(v_0)=r_0$, $r_1(v_0)=1$, $r_{n\ge 2}=0$.

We obtain the required representation (\ref{ZZ})-(\ref{ZZx}) for $Z$ in two steps. At first we
write the following decomposition for the function ${\cal Z}$
\ba
&&{\cal Z}(v,r)={\cal Z}_0(v)+\sum_{m=1} {\Delta r(v,z)^m\over m!} {\cal Z}_m(v) \, ,\n{ffx}\\
&&{\cal Z}_0(v)={\cal Z}(v,r_0(v))\, ,\\
&&{\cal Z}_m(v)=\left. {\pa^m {\cal Z}(v,r)\over \pa r^m}\right|_{r=r_0(v)}\, .\n{fm}
\ea
Then we substitute expression (\ref{rrx}) for $\Delta r(v,x)$ into (\ref{ffx}) and collect the terms of the same power of $z$ in the obtained relation. The result of this algorithmic procedure allows one to find expressions for an arbitrary $Z_n(v)$. Here we just give the first several terms
\ba
Z_0&=&{\cal Z}_0\, ,\nonumber\\
Z_1&=&{\cal Z}_1 r_1 \, ,\n{ZZZZ}\\
Z_2&=&{\cal Z}_1 r_2+{\cal Z}_2 r_1^2 \, ,\nonumber\\
Z_3&=&{\cal Z}_1 r_3+3{\cal Z}_2 r_1 r_2+{\cal Z}_3 r_1^3\, .\nonumber
\ea

\subsection{Beam equations in bracket variables}

Instead of solving the beam equations for $r_n(v)$ it is more convenient to chose other variables, which are connected with $r_n(v)$, but which are more closely related to observables and for which the evolution equations take simpler form.
Let us denote
\ba
&& p(v)=\left. [ r(v,z),z] \right|_{z=0}\, ,\\
&& w(v)=\left. \la r(v,z),z\ra \right|_{z=0}\, ,\\
&& \ve(v)=\left. \{ r(v,z),z\} \right|_{z=0}\, .
\ea
It is easy to show that
\ba
&&p(v)=\ln r_1(v)\, ,\\
&&w(v)={r_2(v)\over r_1(v)}\, ,\\
&&\ve(v)={r_3(v)\over r_1(v)}-{3\over 2} \left( {r_2(v)\over r_1(v)}\right)^2\, .
\ea
We call the functions $p$, $w$ and $\ve$, which characterize deformation of the beam and have nice analytical properties, described above, {\it bracket variables}. The beam equation (\ref{rrr}), written in terms of the bracket variables simplify and take the form
\ba
&&{dr_0\over dv}={\cal Z}_0, \n{Z0}\\
&&{d p\over dv}={\cal Z}_1,\n{Z1}\\
&&{d w\over dv}={\cal Z}_2 e^{p},\n{Z2}\\
&&{d \ve\over dv}={\cal Z}_3 e^{2p}.\n{Z3}
\ea
For our choice of the beam parametrization, one has the following initial data for this system
\be
r_0(v_0)=r_0\hh p(v_0)=w(v_0)=\ve(v_0)=0 .
\ee
It is worth mentioning that the function \eq{Z1}
\be\n{kappa}
{\cal Z}_1=\kappa(v,r)\hh \kappa(v,r)\equiv {1\over 2}{\partial\over \partial r}(\alpha f),
\ee
when taken on the horizon of a static black hole, coincides with its surface gravity.


\section{Calculation of observables}

\subsection{Scheme of the calculations}

For the calculation of the observables $P$, $W$ and ${\cal E}$ at ${\cal I}^+$ one can use either a function $u_-(u_+)$ or a function $u_+(u_-)$ and their derivatives. Correspondingly, one can integrate the system (\ref{Z0})--(\ref{Z3}) over the parameter $v$ either from the future to the past, or from the past to the future. To distinguish these two options we call the first one the {\em top-down} method and the second one the {\em bottom-up} method. Let us describe these methods in more details.

\begin{figure}[tbp]
\centering
\includegraphics[width=6.5cm]{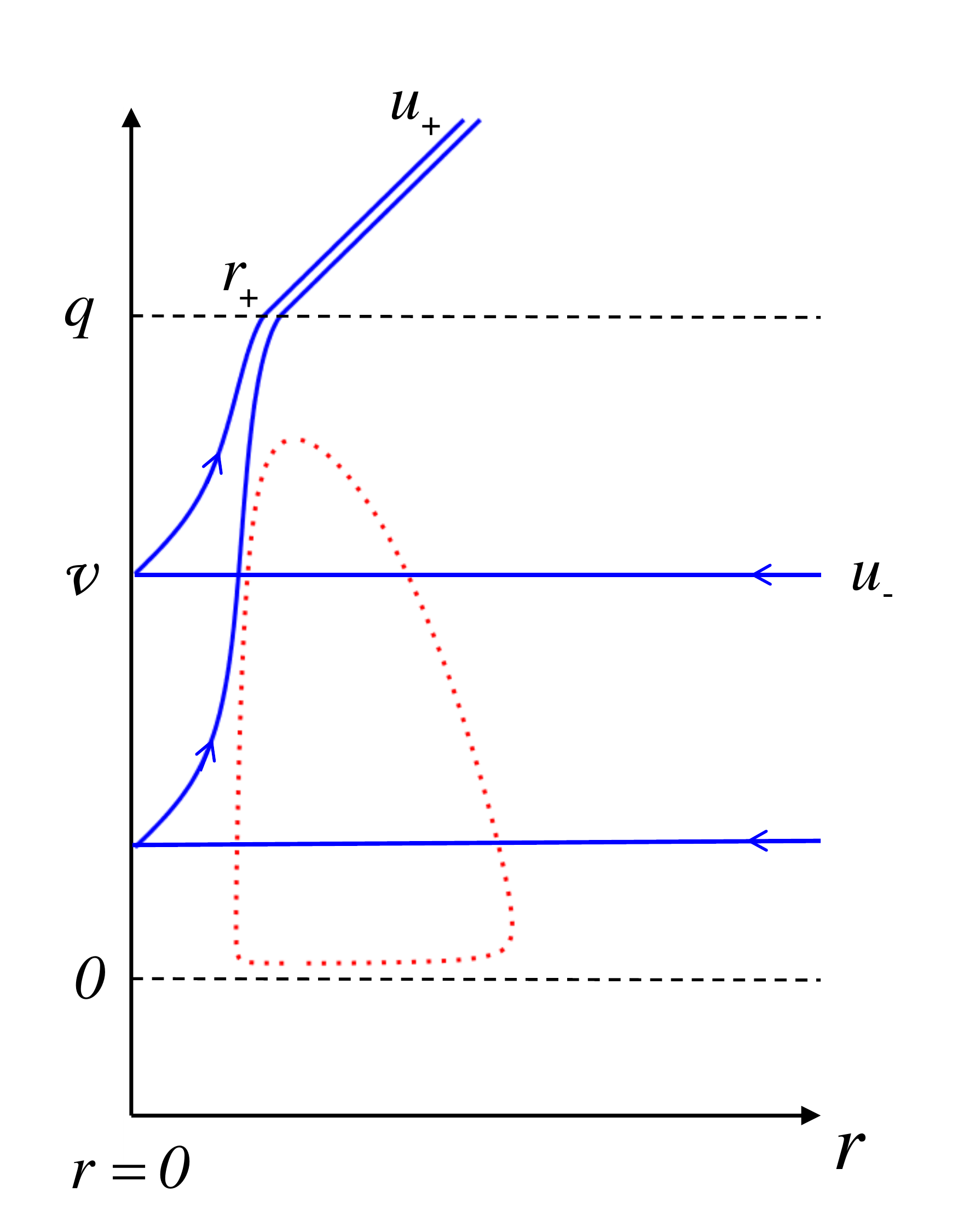}
\caption{ This picture schematically shows
null rays propagating in the nonsingular black hole spacetime \eq{n1}. The dotted curve marks the shape of an apparent horizon. Before the moment $v=0$ and after complete evaporation of the black hole at $v=q$ the background metric is assumed to be flat.
The null ray, which reaches the center $r=0$ at the moment  $v$, leaves the  black hole and crosses  $v=q$ surface at the radius $r=r_{+}$. Then it propagates freely to null infinity.
\label{Fig2}}
\end{figure}

In order to calculate the required observables at ${\cal I}^+$ we proceed as follows. Consider a null ray, which leaves ${\cal I}^-$ at $v=u^-$ and denote by $u^+$ the retarded time, when it arrives to ${\cal I}^+$ (see Figure~\ref{Fig2}). We chose this ray as a fiducial one and consider a beam of null rays in its vicinity. Let $v=q$ be a null surface located in the future spacetime domain, where the spacetime is already flat. Denote by $r^+$ the radius of a point where the fiducial ray crosses $q$-surface. Consider a null ray from the beam, which leaves ${\cal I}^-$ at $u_-=u^-+\Delta v$. Denote by $r_-=x$ the radius where it crosses the $u^-$-surface, and by $r_+=r^+ +y$ the radius of its intersection of $q$-surface. It finally reaches ${\cal I}^+$ at the moment $u_+=u^+ +\Delta u$ of the retarded time. For a fixed fiducial ray the value of the derivatives between the objects in the set $\{u_+,r_+,r_-,u_-\}$
are the same as the value of the corresponding derivatives within the set $\{\Delta u,y,x,\Delta v\}$.

For example, $u_+=q-2 r_+=q-2r^+-2y$, so that
\be
{du_+\over dy}=-2\hh \la u_+,y\ra=0\hh \{u_+,y\}=0\, .
\ee
Relations between $x$ and $u_-$ can be found by local analysis of the ray propagation in the vicinity of the center in the domain close to the $u^-$-surface. These calculations are done in the Appendix. The results are
\be\begin{split}\n{xxuu}
&{dx\over du_-}=-{1\over 2} \alpha_0\, ,\\
&\la x, u_-\ra={\alpha_0'\over \alpha_0}\, ,\\
&\{ x, u_-\}={1\over 2}\alpha_0^2 a_2 +{\alpha_0''\over \alpha_0}-{3\over 2} \left( {\alpha_0'\over \alpha_0}\right)^2\, .
\end{split}\ee
Here $\alpha_0(u_-)$ is the value of the redshift factor $\alpha(v,r)$ at the center, $r=0$, at the moment $v=u_-$ of the advanced time. The dot is a derivative with respect to $u_-$. The function $a_2(u_-)$ is defined as
\be
a_2(v)=\left. {d^2 {\cal Z}(v,r)\over dr^2}\right|_{r=0}\, .
\ee
It is assumed that after performing the required calculations one takes the limit $x=0$ in the objects of the both sides of the relations (\ref{xxuu}).

A gap in the establishing of the relations between objects depending on $u_-$ and $u_+$ can be filled in by similar relations between  the corresponding objects, depending on $x$ and $y$.
This can be done by solving equations (\ref{Z0})--(\ref{Z3}).

\subsection{Bottom-up method}

In the bottom-up method one integrates these equations from $v=u_-$ to the future with the inial conditions
\be
r_0(u_-)=0\hhh p^-(u_-)=w^-(u_-)=\ve^-(u_-)=0\, .
\ee
We use superscript ``$-$`` to remind that the corresponding quantities are specified by their value at $v=u^-$ surface.
The integration is performed forward in time $v$ up to $v=q$. Thus one has
\be\n{pw}
[y,x]=p^-(q)\, ,\ \la y,x\ra= w^-(q)\, ,\ \{ y,x\}= \ve^-(q)\, .
\ee

The map function $u_+=u_+(u_-)$ can be written as the following composition $u_+=u_+(y(x(u_-)))$. Using the chain rules one finds
\ba
&&[u_+,u_-]=p^-(q)+\ln \alpha_0\, ,\nonumber\\
&&\la u_+,u_-\ra=-{1\over 2}\alpha_0 w^-(q)+{\alpha_0'\over\alpha_0}\, ,\\
&&\{ u_+,u_-\}={1\over 4}\alpha_0^2\ve^-(q) +\{x,u_-\}\, .\nonumber
\ea
Here $\{x,u_-\}$ is given by \eq{xxuu}. These relations imply
\ba
[u_-,u_+]~&=&-p^-(q)-\ln \alpha_0\, ,\n{Panom}\\
\la u_-,u_+\ra\, &=& e^{-p^-(q)} \left[ {1\over 2} w^-(q)-{\alpha_0'\over \alpha_0^2}\right]\, ,\n{Wanom}\\
\{ u_-,u_+\}&=&- e^{-2p^-(q)}\!\!\left[{1\over 4}\ve^-(q) +{1\over \alpha_0^2}\{x,u_-\}\!\right]\!.  \n{Eanom}
\ea

\subsection{Top-down method}

In this method one integrates the equations (\ref{Z0})--(\ref{Z3}) backward in time $v$. Namely, one impose the following ``initial'' conditions at $v=q$
\be
r_0(q)=r^+\hhh p^+(q)=w^+(q)=\ve^+(q)=0\, .
\ee
We use superscript ``$+$`` to remind that the corresponding quantities are specified by their value at $v=q$ surface.
The integration is performed backward in time $v$ up to $v=u^-$. Thus one has
\be
{dx\over dy}=\exp[p^+(u_-)]\, ,\ \la x,y\ra= w^+(u_-))\, ,\ \{ x,y\}= \ve^+(u_-)\, .
\ee
Omitting the calculations, we present here the final result
\be\begin{split}\n{uupm}
&\la u_-,u_+\ra=-{1\over 2} w^+(u_-)-e^{p^+(u_-)}{\alpha_0'\over \alpha_0^2}\, ,\\
&\{ u_-,u_+\}={1\over 4} \ve^+(u_-)- { e^{2p^+(u_-)}\over \alpha_0^2} \{x,u_-\}\, .
\end{split}\ee
By comparing (\ref{Wanom}) with (\ref{uupm}) one finds
\ba
&&p^+(u_-)=-p^-(q)\, ,\nonumber\\
&&w^+(u_-)=-w^-(q)e^{-p^-(q)}\, ,\n{recip}\\
&&\ve^+(u_-)=-\ve^-(q)e^{-2p^-(q)}\, .\nonumber
\ea
It is easy to check that these {\it reciprocity conditions} follow directly from the form of the evolution equations (\ref{Z0})--(\ref{Z3}).


\section{Standard model of a nonsingular black hole}

\subsection{Metric}

To specify the model of a nonsingular black hole, one needs to specify the functions $f(v,r)$ and $\alpha(v,r)$ in the metric \eq{n1}. In the literature the Hayward metric \cite{Hayward:2005gi} is a traditional choice describing a model of a static nonsingular black hole
\be\n{Hayward}
f=1-{2M r^2\over r^3 +2M \ell^2}\hh \alpha=1.
\ee
In \cite{Lorenzo,Frolov:2016pav} and our previous paper \cite{Frolov:2016gwl}  the Vaidya type modifications of metric \eq{Hayward}, which also include a nontrivial red shift factor $\alpha(v,r)$, were studied.

Here we consider a very similar metric with
\be\n{a.3}
f=1-{2M(v) r^2\over r^3 +2M(v) \ell^2+\ell^3} .
\ee
In the static case the difference from \eq{Hayward} is an extra $\ell^3$ in the denominator of the function $f$. We assume that $f=1$ outside the interval $v=(0,q)$. It corresponds to $M(v)=0$  for  $v<0$, i.e., before the null matter forming the black hole arrives, and for $v>q$, i.e., after a complete evaporation of the black hole at the moment $v=q$. With this requirement the time dependent version of the Hayward metric \eq{Hayward} has a disadvantage that the curvature tensor is discontinuous at the moments $v=0$ and $v=q$. For our metric \eq{a.3} the evolution of the curvature is smooth at all times. The other nice properties of these geometries \eq{a.3} and \eq{Hayward} are basically the same.

Using the freedom of choice of the constant $\sigma$ in the metric \eq{n1} and a proper rescaling of coordinates, one can always make $\ell=1$. It is also convenient to introduce new notation for the quantity
\be
\mu(v)\equiv 2M(v).
\ee
Then we get
\be\n{aa}
f=1-{\mu(v) r^2\over r^3 +\mu(v)\,+1} .
\ee
We begin with a study of the standard model, when $\alpha=1$.

This metric describes the nonsingular evolving black hole with two apparent horizons: the outer horizon $r_1(v)$ and the inner one $r_2(v)$. Their radii are determined by the condition $f=0$
and can be written explicitly
\be\begin{split}
r_\ins{1}&={1\over 2}\left(\mu+1+\sqrt{\mu^2-2\mu-3}\right),\\
r_\ins{2}&={1\over 2}\left(\mu+1-\sqrt{\mu^2-2\mu-3}\right),
\end{split}\ee
where $\mu=\mu(v)$.
The third root of the cubic equation $f=0$ corresponds to a negative value of radius $r_\ins{0}=-1$ and is not relevant for our study.
The inner and outer apparent horizons exist only when $\mu(v)> 3$.
One can easily calculate the quantity \eq{kappa} on the outer and inner horizons
\be
\kappa_\ins{1,2}=\mu{r_\ins{1,2}(r_\ins{1,2}^3-2\mu-2)\over2(r_\ins{1,2}^3+\mu+1)^2}.
\ee
One can check that, provided horizons exist,  $\kappa_\ins{1}\ge 0$, while $\kappa_\ins{2}\le 0$.
For large masses $\mu\gg 3$ we have
\be\begin{split}
r_\ins{1}&=\mu-{1\over\mu}+O(\mu^{-2}),\\
r_\ins{2}&=1+{1\over\mu}+O(\mu^{-2}),\\
\kappa_\ins{1}&={1\over 2\mu}-{1\over \mu^3}+O(\mu^{-4}),\\
\kappa_\ins{2}&=-1+{5\over 2\mu}+{1\over \mu^3}+O(\mu^{-4}).
\end{split}\ee
For $\mu= 3$ the horizons merge and we obtain an extremal case with $r_\ins{1}=r_\ins{2}=2$ and $\kappa_\ins{1}=\kappa_\ins{2}=0$.

The motion of the incoming radial null rays in this geometry is rather simple. They are described by the equation $v=$const. Such rays pass through the center $r=0$ and then become outgoing (see  Fig.\,\ref{Fig2}). All these null rays can be sorted into three groups.  We denote incoming null rays with $v<0$ as type $I$ rays. During the ingoing stage they propagate in a flat geometry. Similarly after a complete evaporation of the black hole $v>q$ the background spacetime is assumed to be flat again. We denote incoming null rays with $v>q$ as type $III$. These rays propagate in a flat geometry both during the ingoing stage and the outgoing stage, after passing the center $r=0$. The incoming rays with $0\le v\le q$ are of type $II$.

\begin{figure}[tbp]
\centering
\includegraphics[width=6.5cm]{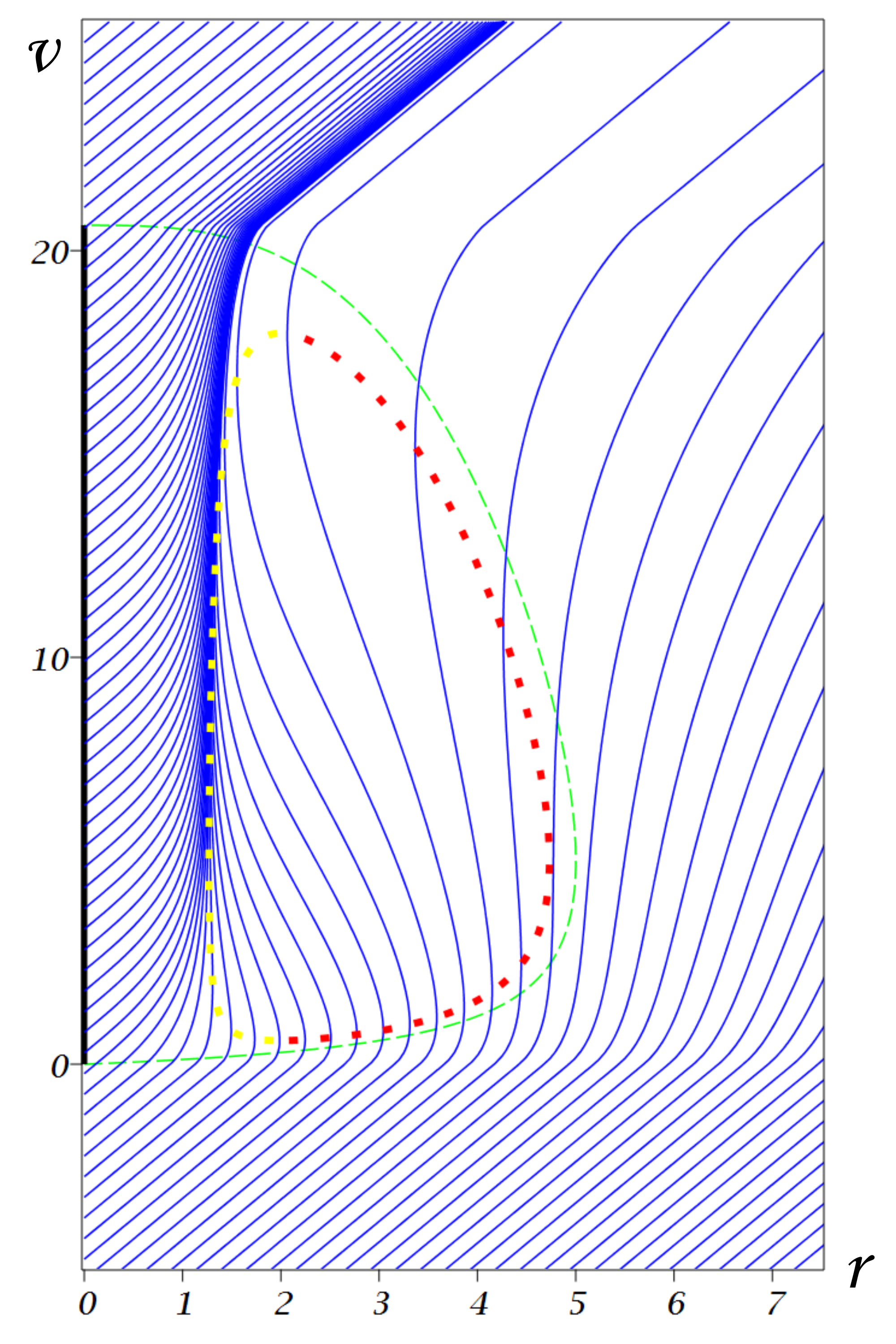}
  \caption{
  This plot shows the outgoing radial null rays $u=\const$ propagating in the nonsingular black hole with $\mu_0\equiv \max \mu(v)=5$. The dotted curve depicts the apparent horizons, inner and outer, while the dashed line marks the function $\mu(v)$.
\label{Fig3}}
\end{figure}

Outgoing null rays in the standard black-hole model are shown in Fig.\,\ref{Fig3}. One can see that type $II$ null rays arriving during the existence of the black hole $0\le v\le q$ pass the center $r=0$ and then are accumulated just below the inner horizon. Some type $I$ null rays  $v\le 0$ after passing the center are also accumulated close to the inner horizon (below and above it). The other type $I$ outgoing null rays do not cross the apparent horizon at all or enter the black hole and get out of it at through the outer horizon. After the black hole evaporation the bunch of accumulated in the vicinity of the inner horizon rays propagate freely to infinity ${\cal I}^+$. They would be visible by a distant observer as a very sharp pulse or radiation. The inner horizon with $\kappa_2<0$ plays the role of an attractor for outgoing rays, while the outer horizon with $\kappa_1>0$ repulses the rays.


\subsection{Energy flux and  the other observables}

In Fig.\,\ref{Fig5}-\ref{Fig7} we present the results of numerical computations of the observables $P,W$, and ${\cal E}$ for the metrics \eq{n1} with $f$ given by \eq{a.3} and $\alpha=1$. We call this metric a standard model of a nonsingular black hole.
Qualitative properties of the quantum particle creation are robust and insensitive to the particular choice of the profile of the mass function $\mu(v)$, which describes the model of creation and subsequent evaporation of the black hole. For our illustrations we chose a smooth function $\mu(v)$. At $v=0$  the function $\mu(v)$ begins with a linear growth of the mass, describing creation of a black hole by collapsing null matter, a short smooth transitional period, and a subsequent long period of evaporation. The duration of the creation of the black hole together with a transitional period are described by a parameter $\tau$ which is assumed to be $\sim \mu_0\equiv \max\mu(v)$. Numerical computations show that quantum particle creation during the short initial stage does not depend much on the details of this stage. Most of the particles are created during the long stage of evaporation of the black hole and in the black hole interior.

\begin{figure}[tbp]
\centering
\includegraphics[width=7.5cm]{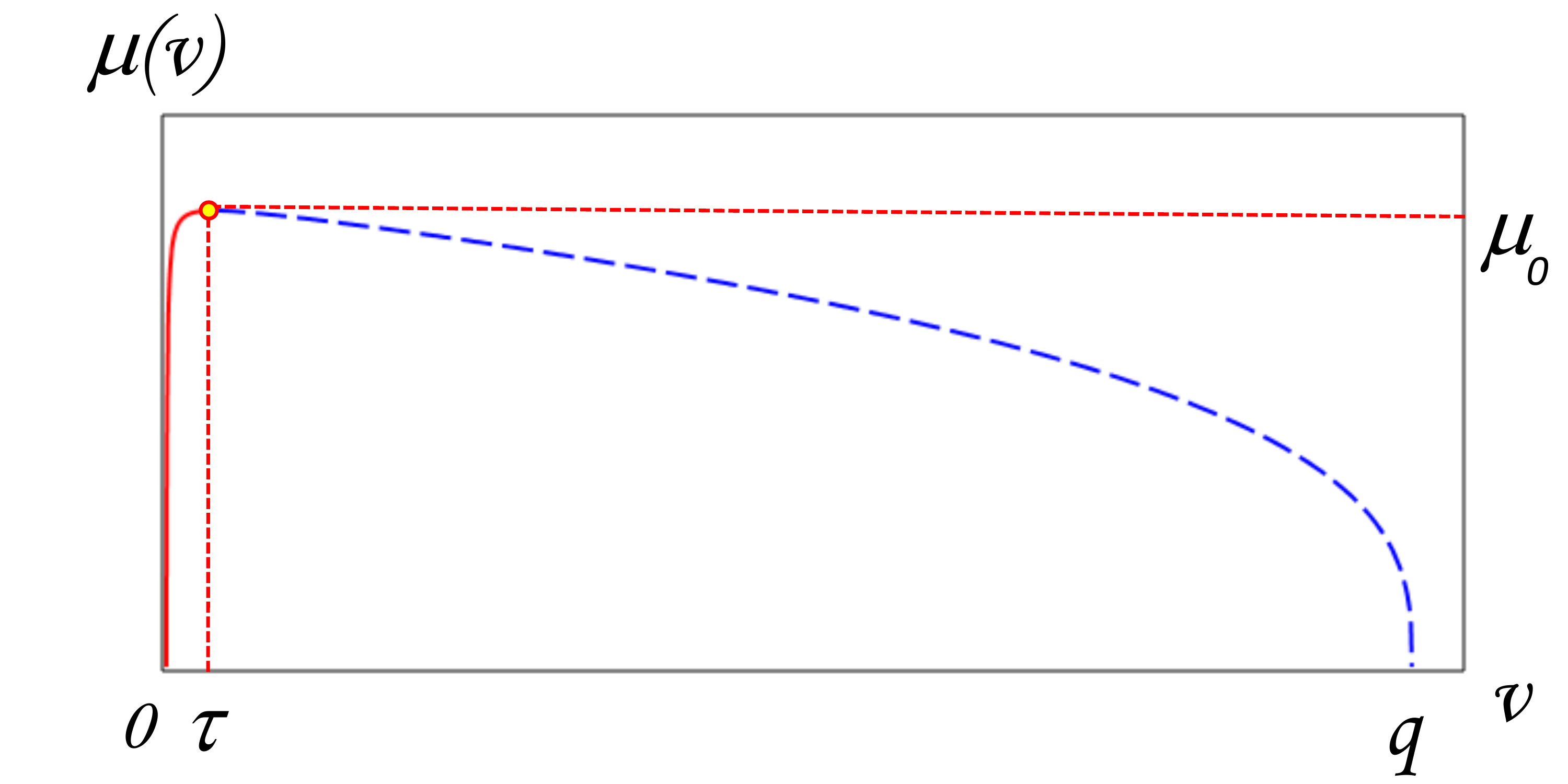}
  \caption{
  This plot, corresponding to $\mu_0=5,~\tau=5,~C=1$,  schematically illustrates a typical shape of the mass function $\mu(v)$. The dashed line describes the stage of the Hawking evaporation of the black hole.
\label{Fig4}}
\end{figure}

\begin{figure}[tbp]
\centering
\includegraphics[width=8cm]{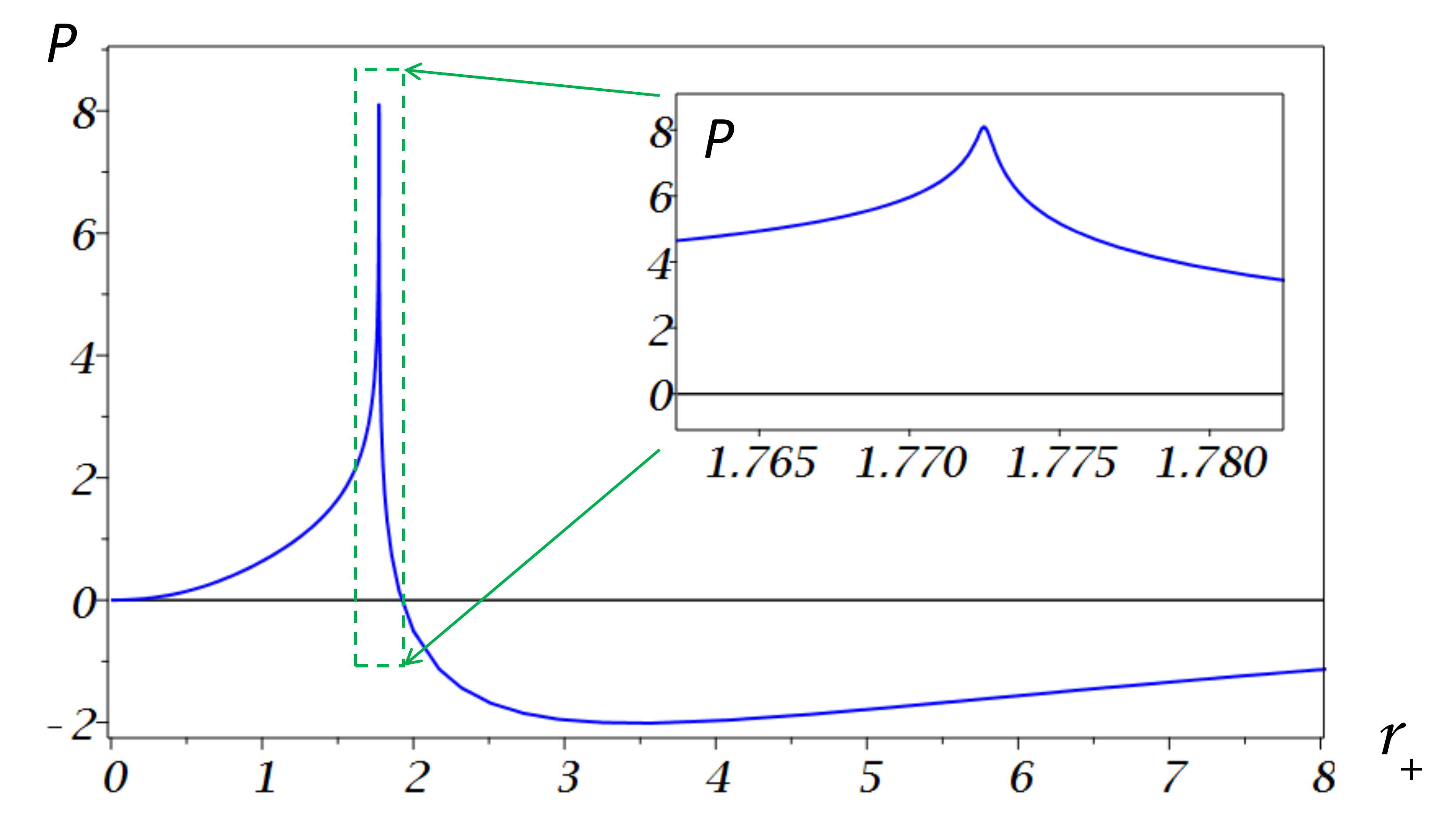}
  \caption{
  This plot shows $P$ as the function of $r_+$. The mass parameter $\mu_0=5$. The sharp peak at the maximum is, in fact, a smooth function.
\label{Fig5}}
\end{figure}

\begin{figure}[tbp]
\centering
\includegraphics[width=8cm]{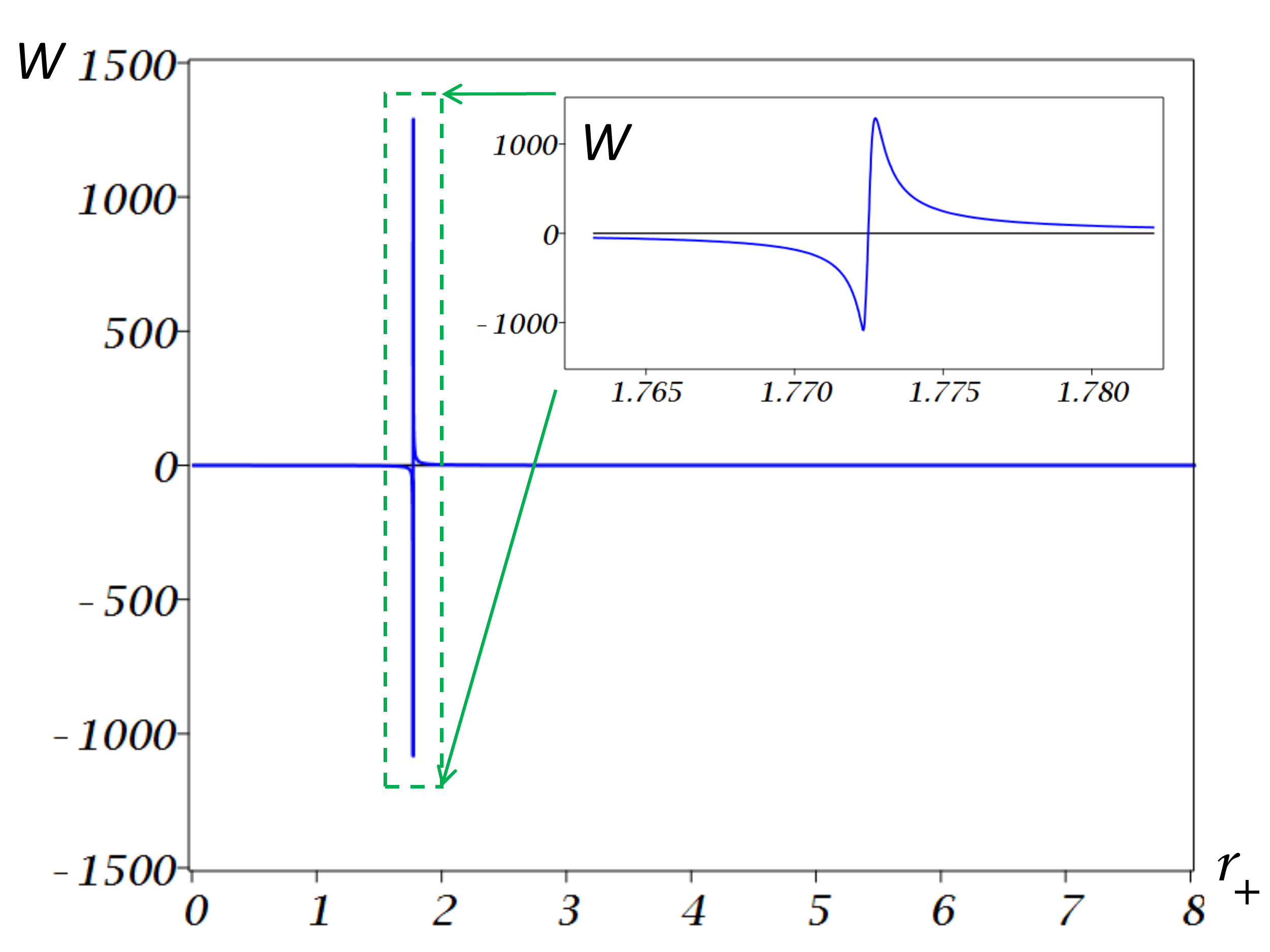}
  \caption{
  This plot shows $W$ as the function of $r_+$. The mass parameter $\mu_0=5$. The sharp peaks concentrated around the position of the inner horizon is, in fact, a smooth curve.
\label{Fig6}}
\end{figure}

\begin{figure}[tbp]
\centering
\includegraphics[width=8.5cm]{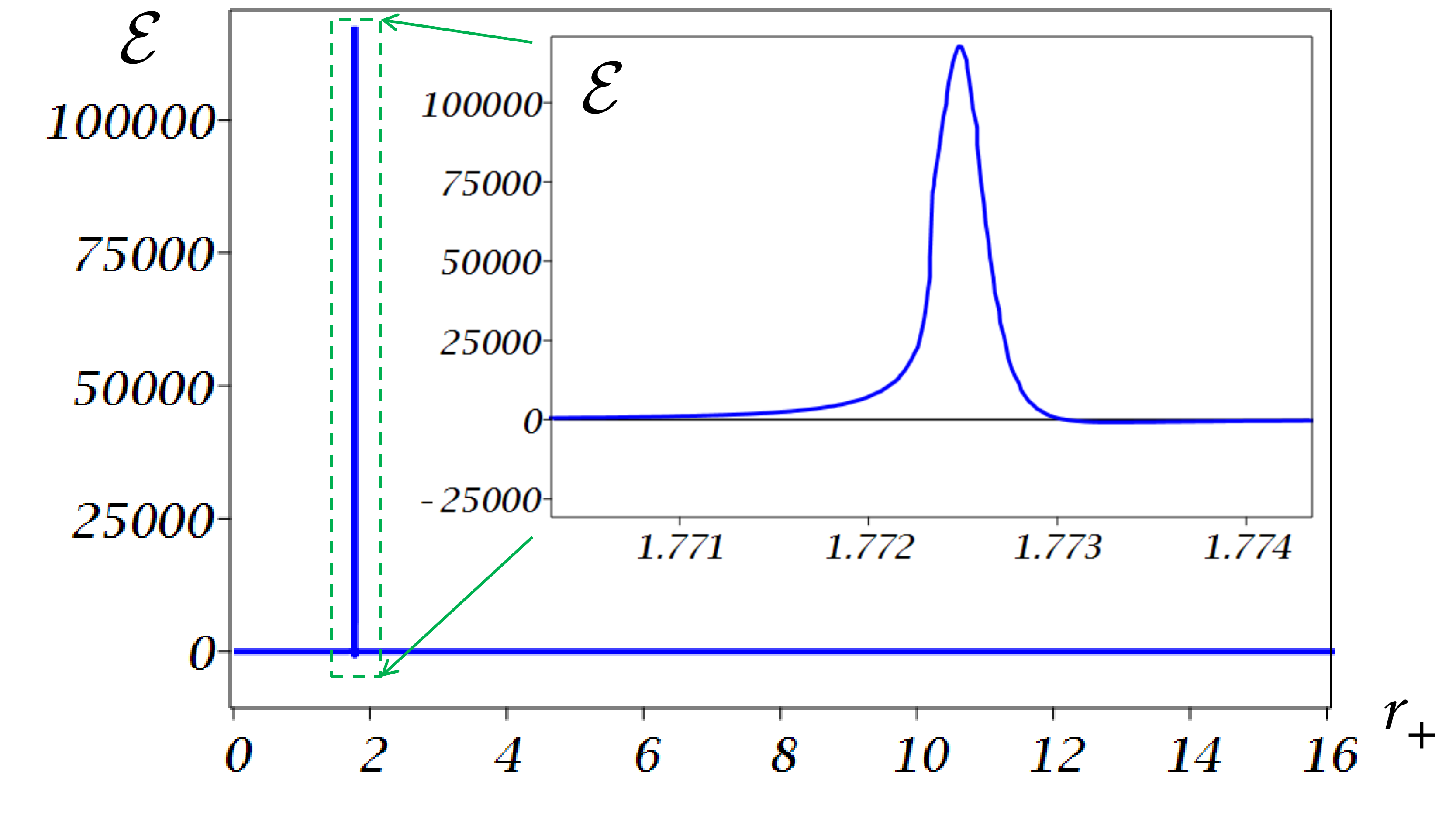}
  \caption{
  This plot shows ${\cal E}$ as the function of $r_+$. The mass parameter $\mu_0=5$. The huge peak of radiation corresponds to rays moving along  the inner horizon. The  zoomed in profile of the peak we put into the box inside the plot.
\label{Fig7}}
\end{figure}

We chose the dynamics of the metric during the evaporation stage qualitatively describing the evaporation of the $4D$ black hole via Hawking radiation
\be
\dot{M}(v)\sim - {N\over 1920\pi}{l_\ins{Planck}^2\over\ell^2}{1\over M(v)^2}
\ee
Here $N$ is the number of distinct polarizations of particles and $M(v)=\mu(v)/2$ is the dimensionless mass of the black hole. Because quantum fluctuations of the metric itself are assumed to be small in the adopted quasiclassical treatment of the geometry, the scale parameter $\ell$ is to be taken larger than $l_\ins{Planck}$. This approximate law of evolution of the black hole mass leads to the total lifetime of the black hole
\be
q=(C\mu_0)^3+\tau\hh
C\sim\left({640\pi\over N}\right)^{1/3}\left({\ell\over l_\ins{Planck}}\right)^{2/3}.
\ee
We choose a function $\mu(v)$ satisfying these conditions,
which is parameterized by two parameters $\mu_0$ and $\tau$. Here $\tau$ is the moment, when the mass of the black hole reaches its maximum $M(\tau)=\mu_0/2$ and the stage of Hawking evaporation begins.  The black hole disappears at the moment $v=q$. A convenient choice (see Fig.\,\ref{Fig4}) of the function $\mu(v)$ reads
\be\n{muv}
\mu(v)=\begin{cases}
\displaystyle{{2v(q-\tau)(q-v)^{1/3}\over C\big(3(q-\tau)v-\tau v+\tau^2\big)}}& 0\le v\le q,\\
0 &  v<0,~~\mbox{and}~~ v>q.
\end{cases}
\ee
This function has a maximum at $v=\tau$, where $\mu(\tau)=\mu_0$.
Note that overall picture of quantum particle creation depends mostly on the asymptotic properties of the function $\mu(v)$ rather than on variations of its shape. Computations using \eq{muv} reveal properties valid for all generic evaporating black hole models.

In order to illustrate the dynamics and quantum effects inside and outside the evolving black hole, in the subsequent numerical computations we fix the parameters of the ansatz \eq{muv} $\mu_0=5$, $\tau=\mu_0$, $C=0.5$.  That is we choose the nonsingular black hole of small mass and the total number of distinct particles polarizations $N$  to be rather large.  In a more realistic case $\mu_0\gg 1$ and $C\gg 1$  and, therefore, $q$ would be very large. Numerical computations can be done anyway, but it would be problematic to present the results graphically. Keeping in mind that qualitative effects look very much alike in all the cases, for the sake of illustration we provide the results only in the case of a small mass nonsingular black hole.

\begin{figure}[tbp]
\centering
\includegraphics[width=8.5cm]{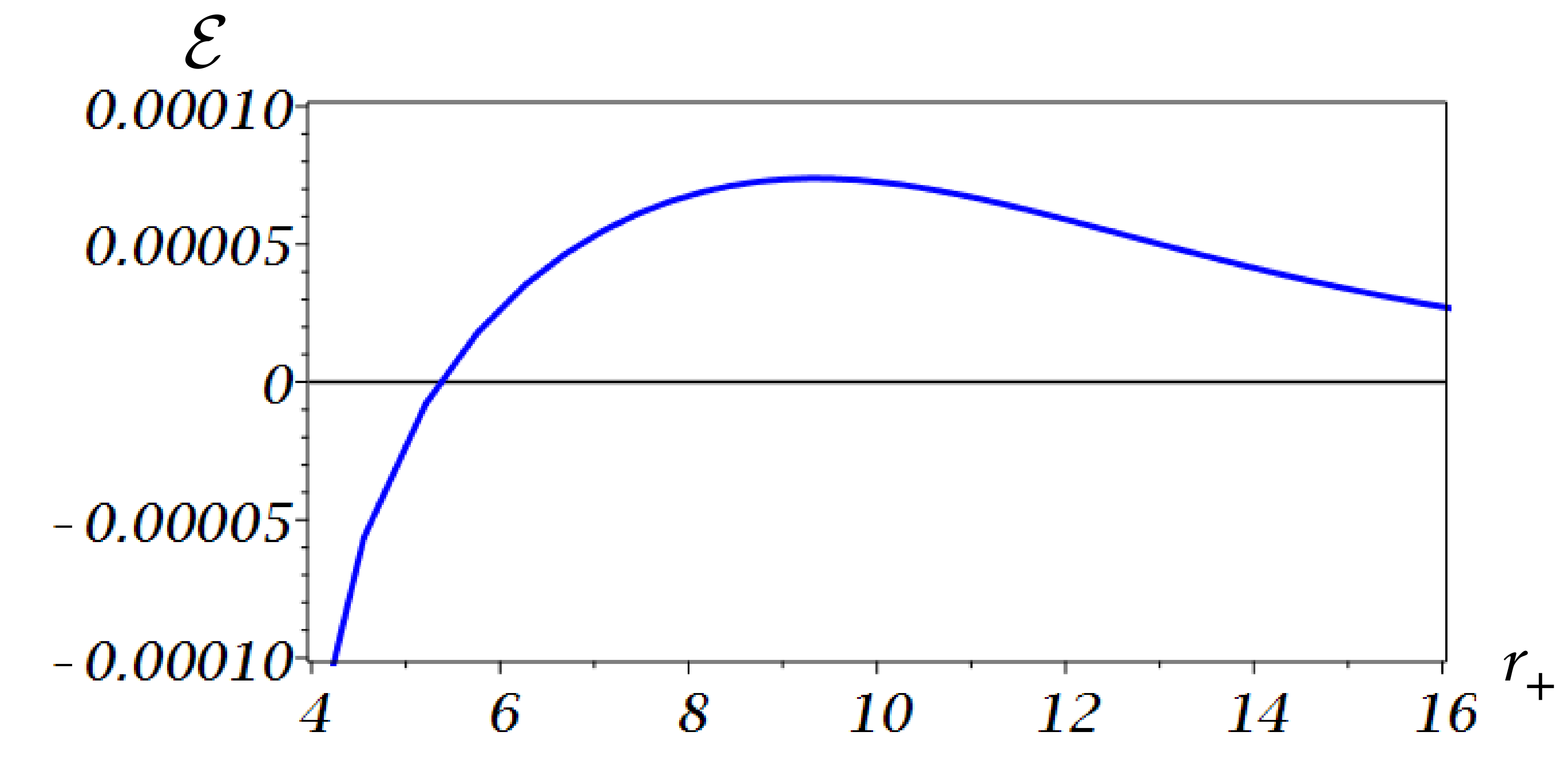}
  \caption{
 The flux of the Hawking radiation ${\cal E}$ at the moment $v=q$ for $r_+>4$. The mass parameter $\mu_0=5$.
\label{Fig8}}
\end{figure}

In Fig.\,\ref{Fig5} we draw the function $P$ (the logarithm of the gain function)  (see \eq{P}, \eq{P1}, \eq{P2}), for the case when the mass parameter $\mu_0=5$. This function has a peak in the vicinity of the inner horizon. For large masses the peak becomes higher and more narrow.
The function W (see \eq{W1},\eq{W2}), which describes the density of out-going null rays, is depicted in Fig.\ref{Fig6}.
The energy flux of massless particles (see \eq{EEEE}, \eq{EESa}, \eq{E2}) is depicted in Fig.\,\ref{Fig7}. The energy flux of particles emitted at the moment $v=q$ from radii $r_+\sim\mu$ and larger describes the Hawking radiation of the evolving black hole. This part of the plot is given in Fig.\,\ref{Fig8}.

Thus, Fig.\,\ref{Fig5}-\ref{Fig8} give the general picture of the quantum radiation as seen by a distant observer at future null infinity. Recall that $u_+=q-2r_+$. Therefore, at the beginning the observer does not see anything. Then Hawking radiation Fig.\,\ref{Fig8} arrives at about $u_+\approx-2\mu$. It lasts with close to constant amplitude (in the case of large masses) till $u_+\approx q-2\mu$. Then the energy fluxes fluctuate considerably and temporarily even may become negative. These fluxes reflect contribution of outgoing particles created inside the black hole between the inner and the outer horizons, and which have been released later after the disappearance of the black hole. The energy flux reaches its maximum (see Fig.\,\ref{Fig7}) at the moment $u_+\approx q-2\rho$,  where
$\rho$ is the radius reached by the outgoing null ray emitted at $r=0,~v=0$ at the moment $v=q$. This pulse of radiation is related with the particles focused at and accelerated near the inner horizon. The amplitude of this peak of radiation is huge and, though its width is exponentially small, the total energy emitted in this
pulse is still exponentially large. Later the flux decreases and eventually vanishes at the moment $u_+=q$.


\section{Modified model}

\subsection{Energy flux and  the other observables}

\begin{figure}[tbp]
\centering
\includegraphics[width=6.5cm]{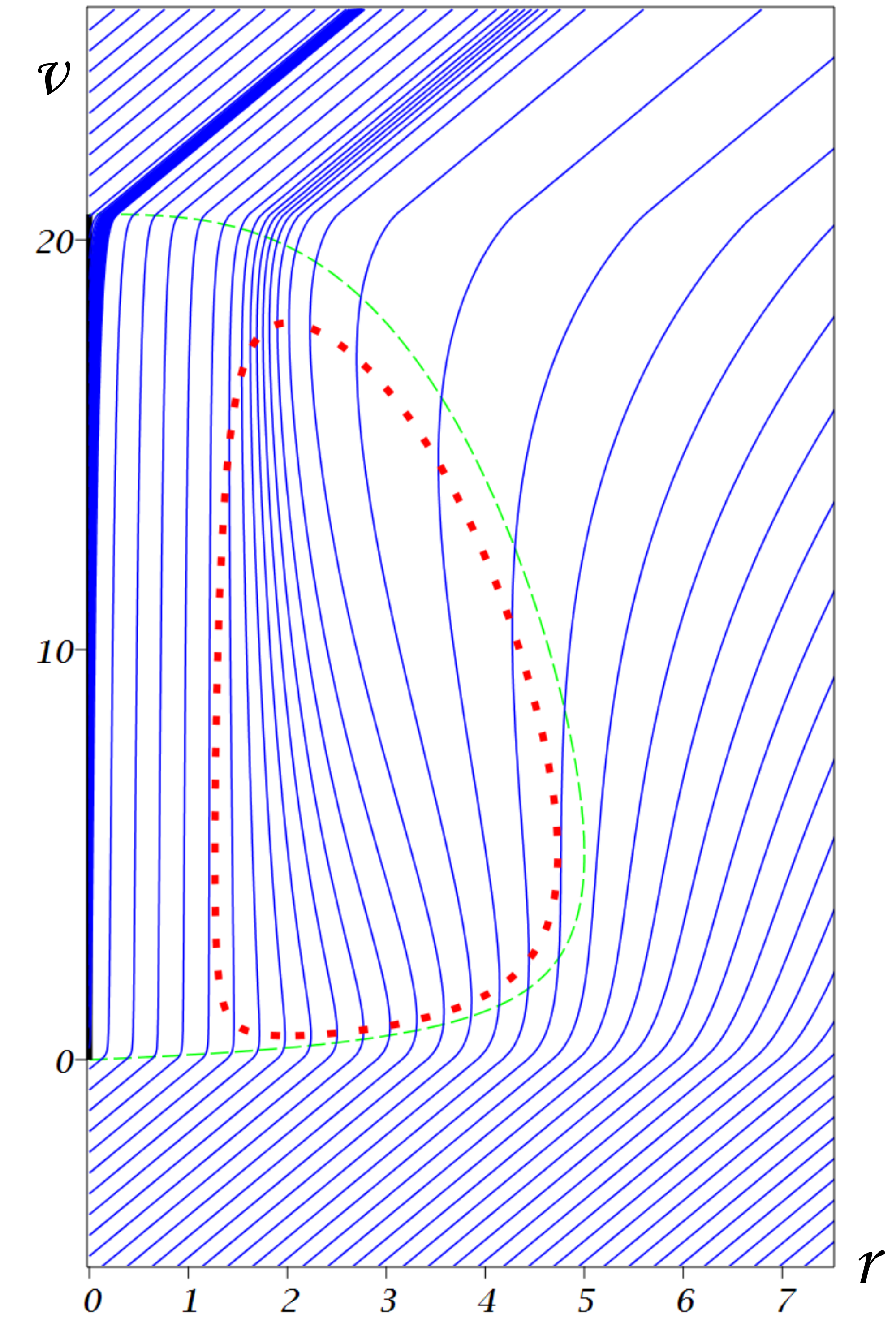}
  \caption{
  This plot shows the outgoing radial null rays $u=\mbox{const}$ propagating in the $\alpha$-modified nonsingular black hole with $\mu_0=5$ and $\alpha=(1+r^5)/(1+r^5+\mu^3)$.
\label{Fig9}}
\end{figure}

Now consider the observables $P,W$, and ${\cal E}$ for the metrics \eq{n1} with
$f$ given by \eq{a.3} and
\be
\alpha={1+r^5\over 1+r^5+\mu^3}.
\ee
We call this metric a modified model of a nonsingular black hole. Evidently, the function $\alpha(v,r)=1$ in the domains, where $\mu(v)=0$, and $\alpha(v,r)\to 1$ at large radii. For large masses $\mu(v)\gg 1$ the role of $\alpha$ at the outer horizon is negligibly small, while in the center and  at the inner horizon it leads to a considerable redshift factor $\sim\mu(v)^{-3}$.
Because of this property time freezes inside the domain near the center of the black hole and the effective (negative) surface gravity (see \eq{kappa12}, \eq{kappa}) of the inner horizon $\kappa_\ins{2}$ is significantly reduced.
As the result the rate of the blue shift of created quanta in the vicinity of inner horizon is suppressed in comparison to the standard model.

In Fig.\,\ref{Fig9} we present the results of numerical computations of the outgoing null trajectories for the modified metrics \eq{n1}. We chose again the mass parameter $\mu_0=5$ for illustration of qualitative properties of the model. One can see that above the outer horizon the picture is qualitatively the same as in the standard model Fig.\,\ref{Fig3}. Inside the black hole the propagation of rays in the modified model and in the standard one differs. There are two peaks of radiation in the modified case. One bunch of null rays is mostly concentrated above the inner horizon, at the finite distance of the order of $\ell$ from the horizon. There is also another domain of concentration of null rays. It is located near the center of the black hole. Its origin is clearly related to the redshift factor $\alpha$ near $r=0$. Because time is freezed there, all ingoing null rays of type $II$ pass the center and then become outgoing rays, which very slowly drift outwards in the vicinity of the center till the complete evaporation of the black hole. Of course, for a very long living black hole (large $q$) these null rays eventually approach the inner horizon from below and merge with the type $II$ bunch of rays to form one pulse of radiation near the inner horizon similar to that of the standard model case. However, for the chosen  $\alpha$ and the mass parameter $\mu_0$ this does not happen.

Numerical computations demonstrate that location of the maximal density of null rays is strongly correlated with the regions, where the energy fluxes are the strongest. For the larger mass parameter $\mu_0$ the picture is qualitatively the same, but the black hole lives much longer $\sim \mu_0^3$ and both peaks of concentration of null rays become even more pronounced.

\begin{figure}[tbp]
\centering
\includegraphics[width=8.5cm]{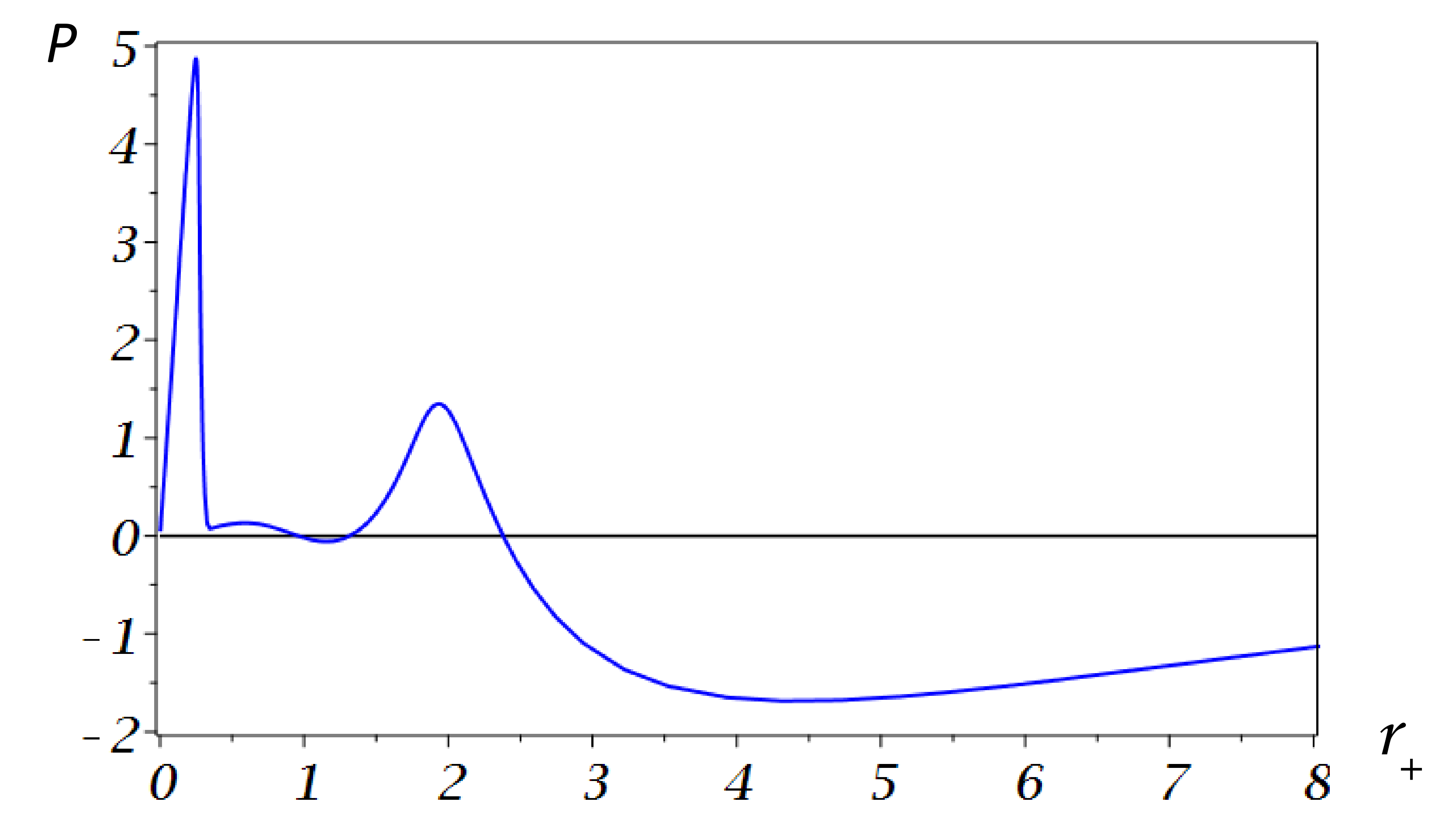}
  \caption{
  This plot shows $P$ as the function of $r_+$ in the modified model. The mass parameter $\mu_0=5$. The high peak on the left at $r\approx 0.27$ is due to the anomalous contribution $-\ln\alpha_0$ in \eq{Panom}.
\label{Fig10}}
\end{figure}

Fig.\,\ref{Fig10} shows the function $P$ (the logarithm of the gain function)  (see \eq{P},\eq{P1},\eq{P2}), for the modified model, when the mass parameter $\mu_0=5$ and $\tau=\mu_0$. This function has two major peaks.  The first one is in the vicinity $r=0$. Its comes from the contribution of the local (anomalous) term $-\ln\alpha_0$ in \eq{Panom}. The second peak originates from $p^-(q)$ in \eq{Panom}. For larger masses this peak becomes more narrow. Both peaks become much higher for large masses. All observables are almost insensitive to the collapse time $\tau$ of null matter forming the black hole, as soon as it is reasonably short. We chose $\tau=\mu_0$ for all plots.

\begin{figure}[tbp]
\centering
\includegraphics[width=8cm]{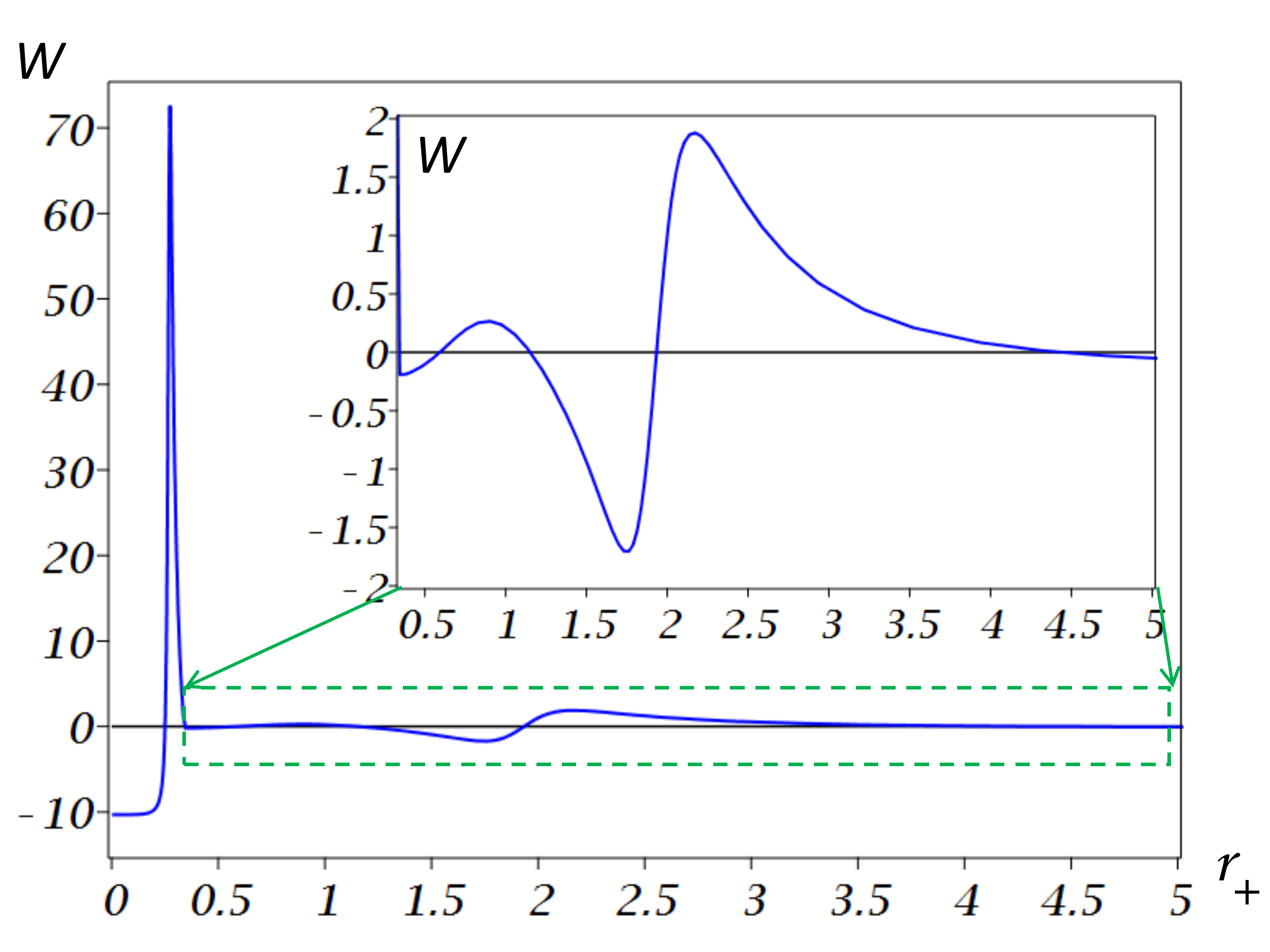}
  \caption{
  This plot shows $W$ as the function of $r_+$ in the modified model. The mass parameter $\mu_0=5$. The high peak on the left at $r\approx 0.27 $ is due to the anomalous contribution in \eq{Wanom}.
\label{Fig11}}
\end{figure}

Fig.\,\ref{Fig11} shows the density of trajectories function $W$   (see \eq{W1},\eq{W2}), for the modified model with the mass parameter $\mu_0=5$.

\begin{figure}[tbp]
\centering
\includegraphics[width=8.5cm]{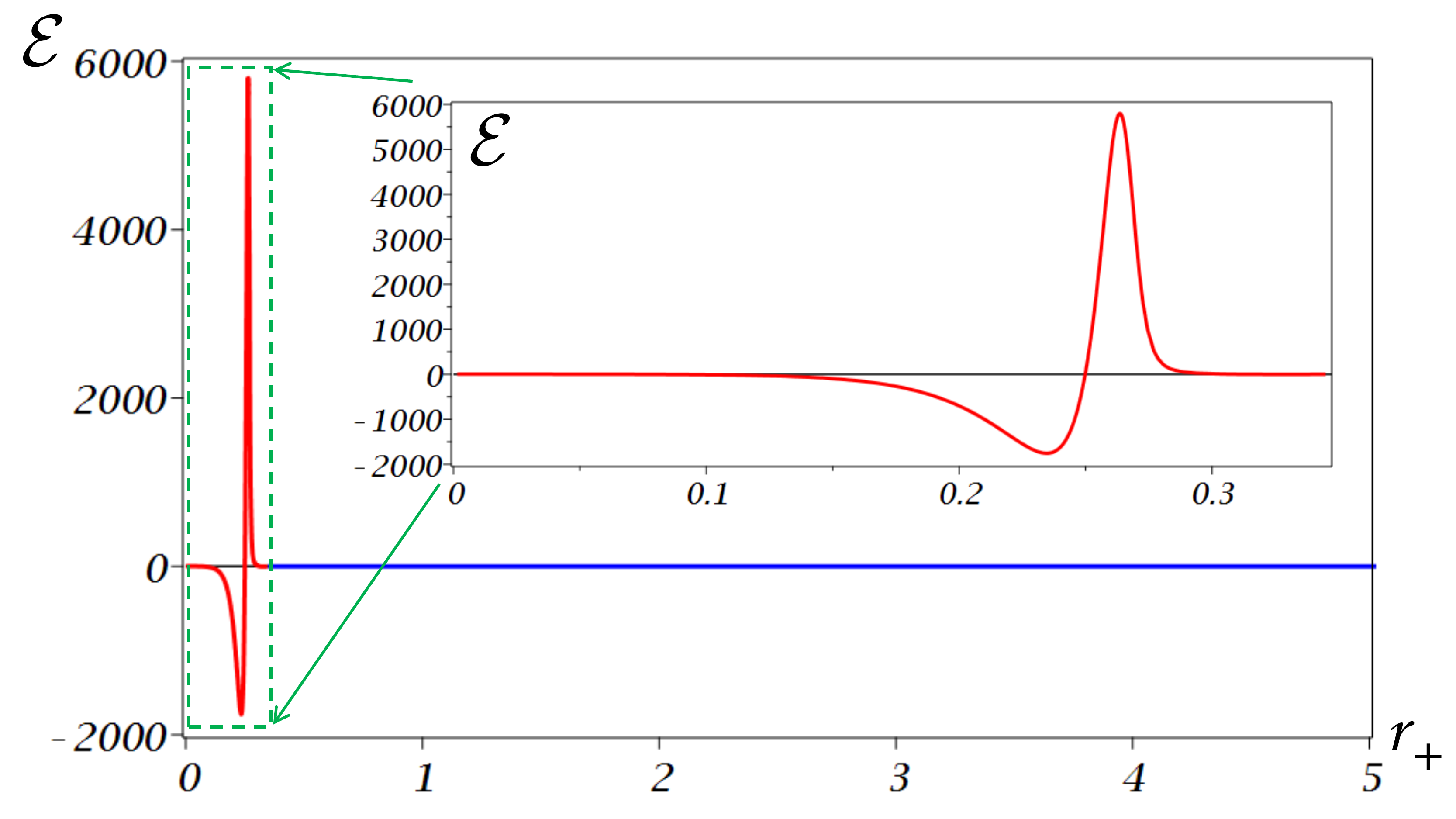}
  \caption{
  This plot shows ${\cal E}$ as the function of $r_+$ in the modified model. The mass parameter $\mu_0=5$. The high peak on the left at $r\approx 0.27 $ is due to the anomalous contribution in \eq{Eanom}. Inside the box we singled out the flux in the range $0\le r_+\le \rho$. For the chosen parameters $\rho\approx 0.3464$.
\label{Fig12}}
\end{figure}

\begin{figure}[tbp]
\centering
\includegraphics[width=8.5cm]{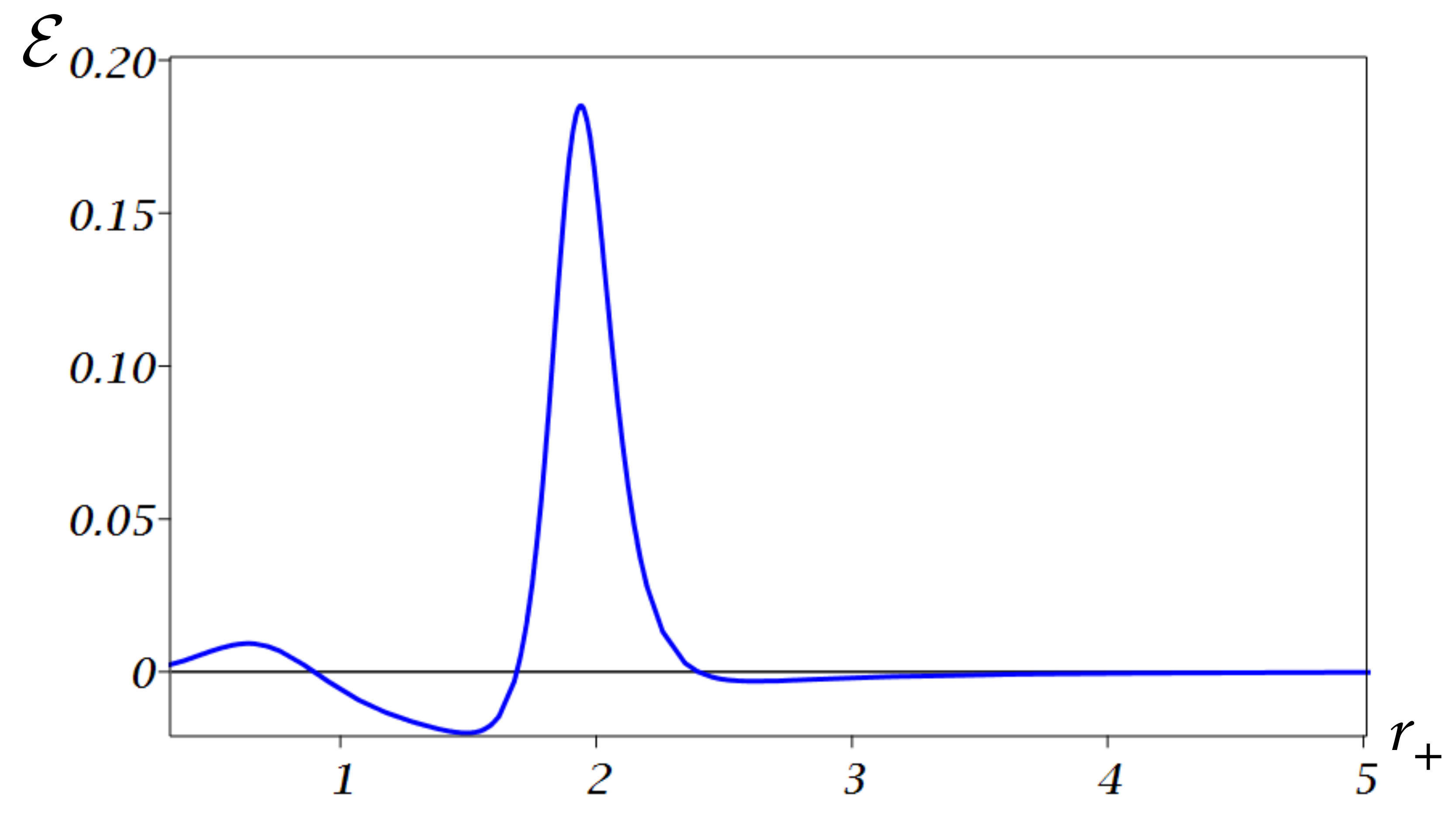}
  \caption{
  This plot shows ${\cal E}$ in the modified model at radii in the interval $0.3464<r_+<5$. The mass parameter $\mu_0=5$.
\label{Fig13}}
\end{figure}

\begin{figure}[tbp]
\centering
\includegraphics[width=8.5cm]{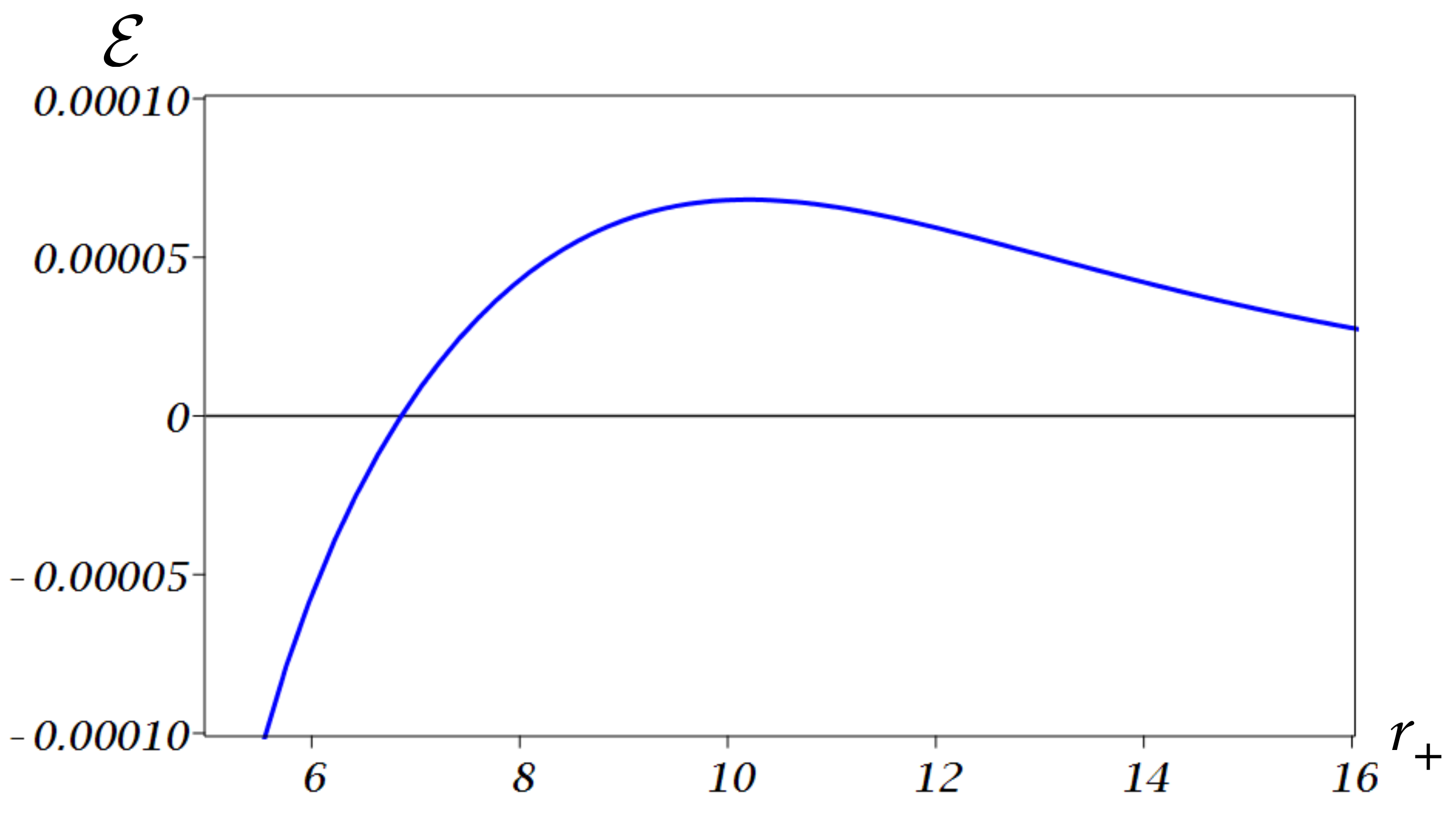}
  \caption{
  The flux of the Hawking radiation ${\cal E}$ at the radii above the outer horizon of the modified model of the evaporating nonsingular black hole. The mass parameter $\mu_0=5$.
\label{Fig14}}
\end{figure}

Figs.\,\ref{Fig12}-\ref{Fig14} depict the energy flux function ${\cal E}$   (see \eq{E2},\eq{Eanom}), for the modified model, when the mass parameter $\mu_0=5$.  One can see that the strongest flux comes from the internal region $r_+\in[0,\rho]$, where
$\rho$ is the radius reached at the moment $v=q$ by the outgoing null ray emitted at $r=0,~v=0$. The profile of the flux is placed in the box inside the Fig.\,\ref{Fig12}. The main contribution to this spike of energy is due to the contribution of the term $\{x,u_-\}/\alpha_0^2$ in \eq{Eanom}. The term $ \ve^-(q)$  in \eq{Eanom} is responsible for the second peak (see Fig.\,\ref{Fig13}) in the energy flux. This peak is of much lesser amplitude but it is still much stronger than the Hawking flux Fig.\,\ref{Fig14} formed in the region above the outer horizon. Because black hole is evaporating, the Hawking flux also changes with time and, evidently, it lasts about the lifetime of the black hole $q$. For large $q$ the Hawking flux is almost constant during the existence of the black hole.

These qualitative properties of the energy flux of quantum radiation are very robust and are model independent. For the sandwich black hole model  \cite{Frolov:2016gwl}, where black hole is turned on and turned of sharply, the particle creation and amplification effects are almost the same. One can see that the energy fluxes are not always positive. This is not surprising  that  because of the quantum nature of the particle creation. Nevertheless the integral flux over the time is to be positive.


\section{Summary and Discussions}

In this paper we studied quantum radiation of massless scalar field from a spherically symmetric non-singular black hole. In the adopted 2D approximation the information concerning the energy flux at infinity is encoded in a function $u_+(u_-)$, which establishes the relation between the retarded  time $u_+$ at the future null infinity and $u_-$, the advanced time at the past null infinity. The corresponding map is provided by radial null rays, propagating in the black hole geometry. It is a quite easy problem to find the function  $u_+(u_-)$ numerically. However, the expression for the quantum energy flux contains the derivatives of this map function up to the third order. Moreover, this function changes very fast near the inner horizon. For this reason we developed a  new method for calculation of classical and quantum observables at ${\cal I}^+$. A starting point of his method is study a beam of outgoing null rays in the vicinity of a chosen (fiducial) radial null ray, connecting points of ${\cal I}^-$ and ${\cal I}^+$. Such rays are parameterised by two variables $(v,x)$. The second of this variables "enumerates" the rays and is constant for a given ray, while the advanced tim $v$ is an affine parameter along such a ray. We found such special combinations of $x$-derivatives of the ray from the beam ("bracket variables") in which the equation of propagation take a simple form. As a result, a function $u_+=u_+(u_-)$ for the fiducial ray, as well as the derivatives of this function up to a given order can be obtained by solving rather simple set of ordinary differential equations.

The expression for the observables on ${\cal I}^+$ can be found either by integration of this set of equations from the past to the future ("bottom-up" method), or by integration of these equations backward in time ("top-down" method). In our calculations we used the first method, which has special advantages. Namely, the expressions for energy flux \eq{EESa} and \eq{Eanom} have the following schematic structure
\be\n{enden}
{\cal E}\sim \beta^2 [A+B]\, .
\ee
The term $B=\alpha_0^{-2}\{ x,u_-\}$ can be interpreted as a contribution to energy flux by particle created in the modes propagating from ${\cal I}^-$ to the center. This is a result of non-adiabaticity of the redshift factor and this contribution vanishes for $\alpha=1$. The term $A\sim \varepsilon^-$ can be interpreted as a contribution of the particles created during the propagation of the outgoing modes from the center to ${\cal I}^+$. The gain function $\beta$ describes the energy amplification effect. Since amplification of the energy is accompanies by the focusing of the null ray beam, the expression for the energy density flux, (\ref{enden}), contains square of the gain function parameter. For the rays propagating in the vicinity of the inner part of the apparent horizon this parameter is large. However, the basic equations of the adopted "bottom-up" method contain not $\beta$ itself, but its logarithm. This provides another advantage for the numerical  calculations.

Results of the calculations confirm the previous results for the quantum radiation, obtained in the so-called sandwich model of a non-singular black hole  \cite{Frolov:2016gwl}.  Namely, for the redshift factor $\alpha=1$ there is an exponentially  large outburst of the energy flux from the inner horizon. This property is directly related to a so-called mass inflation effect \cite{Poisson:1989zz,Poisson:1990eh} (see also \cite{Brown:2011tv} for the more recent study of the mass inflation in the loop gravity black holes). The inner horizon has a negative surface gravity and it works as an attractor of the outgoing null rays. For a (quasi) static case, the advanced time $v$ is the Killing time parameter. The affine parameter $\lambda$ along a null ray close to the inner horizon is related to $v$ as follows $\lambda\sim \exp(-\kappa v)$. The momentum $p^{\mu}=dx^{\mu}/d\lambda$ of outgoing photons exponentially grows, while their energy, $E\sim -p_{v}$, remains constant. This happens, because photon becomes exponentially close to the horizon, and the corresponding gravitational redshift effect compensates the grows of photon frequency. However, this compensation mechanism is broken when the position of the inner horizon moves. As a result, quanta leaving a black hole interior possess such exponentially large non-compensated blueshift.

A special choice of the redshift factor allows one to reduce the flux of the energy calculated for the model with $\alpha=1$. In the model presented in section~V the exponential factor is suppressed. This happens because the surface gravity of the inner horizon is reduced by the factor $\sim \alpha_0$. As a result, the time of the black hole evaporation is not sufficient for generation the exponential regime, and the energy flux decreases. In this regime the main effect is the amplification of the energy in the modes passing through the black hole. This reason of this effect is very simple: during the time of the propagation of the mode in the black hole interior, the mass of the black hole considerably decreases, so that the quanta go away to the infinity from the domain with the gravitational potential smaller, that it was when the corresponding mode enters the black hole.

The  energy flux for the model with $\alpha\ne 1$ it still  large. Partly this is connected with large contribution of the $\alpha$-anomaly. This contribution describes the particle creation in the incoming modes, which becomes large for fast change of the redshift function $\alpha$ during the formation of the black hole. As a result,  the model also violates a self-consistency requirement.
The obtained results indicate that the backreaction of the particles created  in the black hole interior should be properly taken into account.
An interesting problem is a search for self-consistent non-singular black hole models.

\appendix

\section{$u_-\to x$ map}

To find the functions relating $u_-$ and $x$ we proceed as follows.
It is convenient to introduce the center proper time coordinate $\tau$ as follows
\be
\tau=\tau(v)\hh \tau(v)=\int_{u_-}^v dv\, \alpha_0(v)\, .
\ee
Here $\alpha_0(v)$ is the value of the redshift function $\alpha(v,r)$ at the center
\be
\alpha_0(v)=\alpha(v,r=0)\, .
\ee
Let us notice that for such a choice, the parameter $\tau$ vanishes at $v=u_-$ and is negative below this line. We shall us the following notation for the outgoing null rays from the beam in this domain
\be
\rho(\tau,x)=r(v(\tau),x)\, .
\ee

A null ray crosses $v=u_-$ surface at $x$, so that one has $\rho(0,x)=r(u_-,x)$. Being traced backward in time, it  reaches the center $r=0$ at proper time $\tau$. The condition
\be\n{ttx}
\rho (\tau,x)=0\, ,
\ee
establishes the relation between $\tau$ and $x$, which we write in the form
\be\n{tV}
\tau=V(x)\, .
\ee
To establish the relation between $x$ and $u_-$ and to calculate objects $\la x,u_-\ra$ and $\{ x,u_-\}$ we use the following chain of maps
\be\n{chain}
u_- \to \tau \to x \, .
\ee

It is a trivial exercise to show that
\ba
&& \la \tau,u_-\ra={\alpha_0'\over \alpha_0}\, ,\\
&&\{\tau,u_-\}={\alpha_0''\over \alpha_0}-{3\over 2}\left( {\alpha_0'\over \alpha_0}\right)^2\, .
\ea
As earlier, we use the prime to denote a derivative of a function of one variable with respect to its argument. In particular, in the above formulas $\alpha_0'=d\alpha_0/du_-$, etc.

Next, we need to derive an expression for $\{ x,\tau \}$. Using (\ref{yx})  and (\ref{yyxx}) we can write
\ba
&& \la x,\tau\ra=-\la V,x\ra (V')^{-1}\, ,\\
&&\{ x,\tau \}=-\{ V,x \} (V')^{-2}\, .
\ea
To find a function $V(x)$ we need to solve the equation
\be\n{rvx}
\rho(V(x),x)=0\, .
\ee
The derivatives of (\ref{ttx}) give
\ba
&&\dot{\rho} V'+\pa_x \rho=0\, ,\nonumber\\
&&\dot{\rho} V''+\pa_x^2 \rho+\ddot{\rho}(V')^2+2\pa_x\dot{\rho} V'=0\, ,\n{VVVV}\\
&&\dot{\rho} V'''+\pa_x^3\rho +3\ddot{\rho}V' V''+3\pa_x\dot{\rho} V''\nonumber\\
&&\quad\quad+ 3 \pa_x^2 \dot{\rho} V'+3\pa_x \ddot{\rho} (V')^2+\dddot{\rho} (V')^3=0\, .\nonumber
\ea
It should be emphasized that after the corresponding derivatives in these equations are calculated, one must restrict the expressions in the left-hand sides on the line $r=0$.

The function $\rho(\tau,x)$ obeys the equation
\be\n{drr}
\dot{\rho}(\tau,x)= {\cal X}(\tau,\rho)\hh{\cal X}(\tau,\rho)=\alpha_0^{-1}{\cal Z}(v,r)\, .
\ee
The function ${\cal X}$ has the following expansion  near the center
\be\n{ZZZ}
{\cal X}(\tau,\rho)={\alpha f\over 2\alpha_0}={1\over 2}+{1\over 2} a_2(\tau) \rho^2+\ldots\, .
\ee
The dot denotes a derivative over the parameter $\tau$ along the rays, that is for fixed $x$.
Simple calculations give
\ba
&&\ddot{\rho}(\tau,x)=\pa_{\tau} {\cal X}+{\cal X} \pa_{\rho}{\cal X}\, ,\n{z1}\\
&&\dddot{\rho}(\tau,x)=\pa_{\tau}^2 {\cal X}+2 {\cal X}\pa_{\tau} \pa_{\rho}{\cal X}\nonumber\\
&&\quad \quad \quad+{\cal X}^2\pa_{\rho}^2 {\cal X}+\pa_{\tau} {\cal X}\pa_{\rho}{\cal X}+{\cal X}(\pa_{\rho}{\cal X})^2\, ,\n{z2}\\
&&\pa_x\dot{\rho}(\tau,x)= \pa_{\rho}{\cal X} \pa_x \rho\, ,\n{z3}\\
&&\pa_x^2\dot{\rho}(\tau,x)= \pa_{\rho}^2{\cal X} (\pa_x \rho)^2+ \pa_{\rho}{\cal X} \pa_x^2 \rho\, ,\n{z4}\\
&& \pa_x\ddot{\rho}(\tau,x)=\pa_x \rho\left[\pa_{\tau} \pa_{\rho}{\cal X}+
(\pa_{\rho}{\cal X})^2+{\cal X} \pa_{\rho}^2{\cal X} \right]\, .\n{z5}
\ea

Equation (\ref{ZZZ}) implies that at the center $r=0$ one has
\ba
&&{\cal X}={1\over 2}\hhh \pa_{\tau}{\cal X}=0\hhh \pa_{\tau}^2{\cal X}=0\, ,\nonumber\\
&&\pa_{\rho}{\cal X}=0\hhh \pa_{\tau}\pa_{\rho}{\cal X}=0\hhh\pa^2_{\rho}{\cal X}=a_2\, .
\ea
Thus, being calculated at the center, the relations (\ref{z1})--(\ref{z5}) give
\ba
&& \dot{\rho}={1\over 2}\hh\ddot{\rho}=\pa_x\dot{\rho}=0\hh \dddot{\rho}={1\over 4}a_2\, ,\nonumber\\
&&\pa_x^2\dot{\rho}=a_2 (\pa_x\rho)^2\hh
\pa_x\ddot{\rho}={1\over 2}a_2 \pa_x\rho\, .
\ea
By substituting these relations in (\ref{VVVV}) one obtains
\ba
&&V'=-2\pa_x\rho\hh V''=-2\pa_x^2\rho\, ,\nonumber\\
&& V'''=-2 \pa_x^3\rho-4a_2(\pa_x\rho)^3\, .
\ea
Once again, these relations are valid at the center $r=0$.

Let us finally take the limit $\tau\to 0$, which means that the corresponding Schwarzian is calculated at the fiducial ray. At the fiducial ray $v=u_-$ at $x=0$ one has
\be\begin{split}
&V'=-2\hh     V''=0\hh    V'''=4a_2,\\
&\la V,x\ra= 0\hh \{V,x\}=-2 a_2\, .\n{VVxx}
\end{split}\ee
Combining the above results one gets
\ba
&& {dx\over du_-}={\alpha_0\over V'}=-{1\over 2}\alpha_0\, ,\\
&& \la x,u_-\ra={\alpha_0'\over \alpha_0}\, ,\\
&& \{x, u_-\}=\{ \tau, u_- \}+{1\over 2}\alpha_0^2 a_2\, .
\ea

\medskip

\section*{Acknowledgments}

The authors thank the Natural Sciences and Engineering Research Council of Canada and the Killam Trust for their financial support.

%

\end{document}